\documentclass[a4paper,aps,pre,showpacs,twocolumn]{revtex4}

\usepackage{amsmath,amsbsy,amsfonts,amssymb,graphics,epsfig,float,mathrsfs,amsthm,braket}

\usepackage{ifsym}
\usepackage{psfrag}
\usepackage{color}
\usepackage[normalem]{ulem}
\usepackage{multirow}

\newcommand{\tTdry}{\tilde{T}_{\mathrm{dry}}}
\newcommand{\Tpre}{T_{\mathrm{pre}}}
\newcommand{\Tedge}{T_{\mathrm{edge}}}
\newcommand{\dTcap}{\Delta T_{\mathrm{cap}}}
\newcommand{\Tinf}{T_{\infty}}
\newcommand{\Tdry}{T_{\rm dry}}
\newcommand{\Twet}{T_{\rm wet}}
\newcommand{\To}{T_O}

\newcommand{\Ti}{T_I}
\newcommand{\bTi}{\bar{T}_I}
\newcommand{\bTo}{\bar{T}_O}
\newcommand{\bTinf}{\bar{T}_{\infty}}

\newcommand{\barpsi}{\bar{\psi}}
\newcommand{\psiout}{{\psi_{\mathrm{out}}}}

\newcommand{\tpsiin}{{\bar{\bar{\psi}}_{\mathrm{in}}}}
\newcommand{\barpsiin}{\bar{\psi}_{\mathrm{in}}}
\newcommand{\barpsiout}{\bar{\psi}_{\mathrm{out}}}
\newcommand{\srr}{\sigma_{rr}}
\newcommand{\sqq}{\sigma_{\theta\theta}}
\newcommand{\barr}{\bar{r}}

\newcommand{\tauinf}{\bar{T}_{\infty}}
\newcommand{\btauinf}{\tau_\infty}

\newcommand{\glv}{\gamma}
\newcommand{\gsl}{\gamma_{sl}}
\newcommand{\gsv}{\gamma_{sv}}

\newcommand{\tglv}{\tilde{\gamma}}
\newcommand{\epscap}{\epsilon}

\newcommand{\dtheta}{\Delta\theta}
\newcommand{\phiin}{{\phi_{\mathrm{in}}}}
\newcommand{\phiout}{{\phi_{\mathrm{out}}}}

\newcommand{\Usurf}{U_{\mathrm{surf}}}
\newcommand{\Ustrain}{U_{\mathrm{strain}}}
\newcommand{\Ubend}{U_{\mathrm{bend}}}

\newcommand{\lm}{\ell_m}
\newcommand{\last}{\ell_\ast}
\newcommand{\lbc}{\ell_{BC}}
\newcommand{\Li}{L_I}
\newcommand{\Lo}{L_O}
\newcommand{\Rout}{R_{\mathrm{out}}}
\newcommand{\Rin}{R_{\mathrm{drop}}}
\newcommand{\Rcurv}{R_{\mathrm{curv}}}
\newcommand{\rmur}{{u}_r}
\newcommand{\rmurin}{{u}_r^{\mathrm{in}}}
\newcommand{\rmurout}{{ u}_r^{\mathrm{out}}}

\newcommand{\upd}{\mathrm{d}}
\newcommand{\beq}{\begin{equation}}
\newcommand{\eeq}{\end{equation}}

\usepackage{epstopdf}

\begin{document}
\title{Partial wetting of thin solid sheets under tension}

\author{Benny Davidovitch$^1$ and Dominic Vella$^2$}
\affiliation{$^1$Physics Department, University of Massachusetts, Amherst, Massachusetts 01003, USA\\
$^2$Mathematical Institute, University of Oxford, Woodstock Rd, Oxford OX2 6GG, United Kingdom}

\begin{abstract}
We consider the equilibrium of liquid droplets sitting on thin elastic sheets that are subject to a boundary tension and/or are clamped at their edge. We use scaling arguments, together with a detailed analysis based on the F\"{o}ppl-von-K\'{a}rm\'{a}n equations, to show that the presence of the droplet may significantly alter the stress locally if the tension in the dry sheet is weak compared to an intrinsic elasto-capillary tension scale $\glv^{2/3}(Et)^{1/3}$ (with $\glv$ the droplet surface tension, $t$ the sheet thickness and $E$ its Young modulus). Our detailed analysis suggests that some recent experiments may lie in just such a ``non-perturbative" regime. As a result, measurements of the tension in the sheet at the contact line (inferred from the contact angles of the sheet with the liquid--vapour interface) do not necessarily reflect the true tension within the sheet prior to wetting. We discuss various characteristics of this non-perturbative regime. 
\end{abstract}

\maketitle

\section{Introduction}

\subsection{Background}

On rigid, thick solid substrates, partial wetting is governed by the classic Young--Laplace--Dupr\'{e} (YLD) law, which expresses the contact angle, $\theta_Y$, in terms of the surface energies of the three phases that meet at the contact line through the famous equation:
\begin{equation}
\cos\theta_Y = \frac{\Delta\gamma_s}{\gamma}.
\label{eq:LaplaceLaw}
\end{equation}
Here $\gamma\equiv \gamma_{lv}$ is the liquid-vapour surface tension, and $\Delta\gamma_s = \gsv - \gsl$, is the difference between the surface energies of the solid with the surrounding liquid and vapour phases. Importantly, Eq.~(\ref{eq:LaplaceLaw}) is unaffected by the presence (or lack thereof) of a tensile load exerted  on the rigid solid near its surface  --- the partial wetting problem can be described as the minimization of surface energy, $\Usurf$, alone, subject to the constraint that the solid substrate retains its original ({\emph{i.e.}} dry), flat shape.

The burst of applications of elasto-capillary phenomena at micro- and nano-scales has  recently led to a renewed interest in the partial wetting of ultra-thin solid sheets. The sheets used experimentally are typically either glassy (with Young modulus $E\sim1\mathrm{~GPa}$ and thickness $t\sim100\mathrm{~nm}$) or elastomeric (with Young modulus $E\sim1\mathrm{~MPa}$ and thickness $t\sim1\mathrm{~\mu m}$). The stretching modulus of such sheets, $Y=Et$, then typically lies in the range $1\mathrm{~Nm^{-1}}\lesssim Y\lesssim 100\mathrm{~Nm^{-1}}$ and so is  usually much larger than the interfacial tension of a deposited drop $(\gamma\sim0.1\mathrm{~Nm^{-1}})$, i.e.~$\gamma/Y \ll1$. As a result,  under characteristic capillary-induced loads, the sheet may be considered to be  nearly inextensible.

As was noted by Olives \cite{Olives93}, who built upon previous work by Shanahan \cite{Shanahan87}, the partial wetting of a thin solid sheet reflects the simultaneous minimization of three energies: The surface energy, $\Usurf$, as well as the elastic energies $\Ustrain$ and $\Ubend$, associated, respectively, with the anisotropic, non-uniform distribution of strain and the curvature induced in the solid sheet by the presence of the liquid drop. The nontrivial nature of the problem, in comparison with YLD law (\ref{eq:LaplaceLaw}), emanates from the subtle interplay between these three energies in solid sheets that are ``highly-bendable'' yet ``nearly inextensible".  An intimately related complication that has emerged in more recent applications \cite{Nadermann13,Schulman15,Schulman17} is the presence of a uniform isotropic tension, which typically exists in the sheet prior to its wetting by the liquid drop.

\begin{figure}
\centering
\includegraphics[width=0.9\columnwidth]{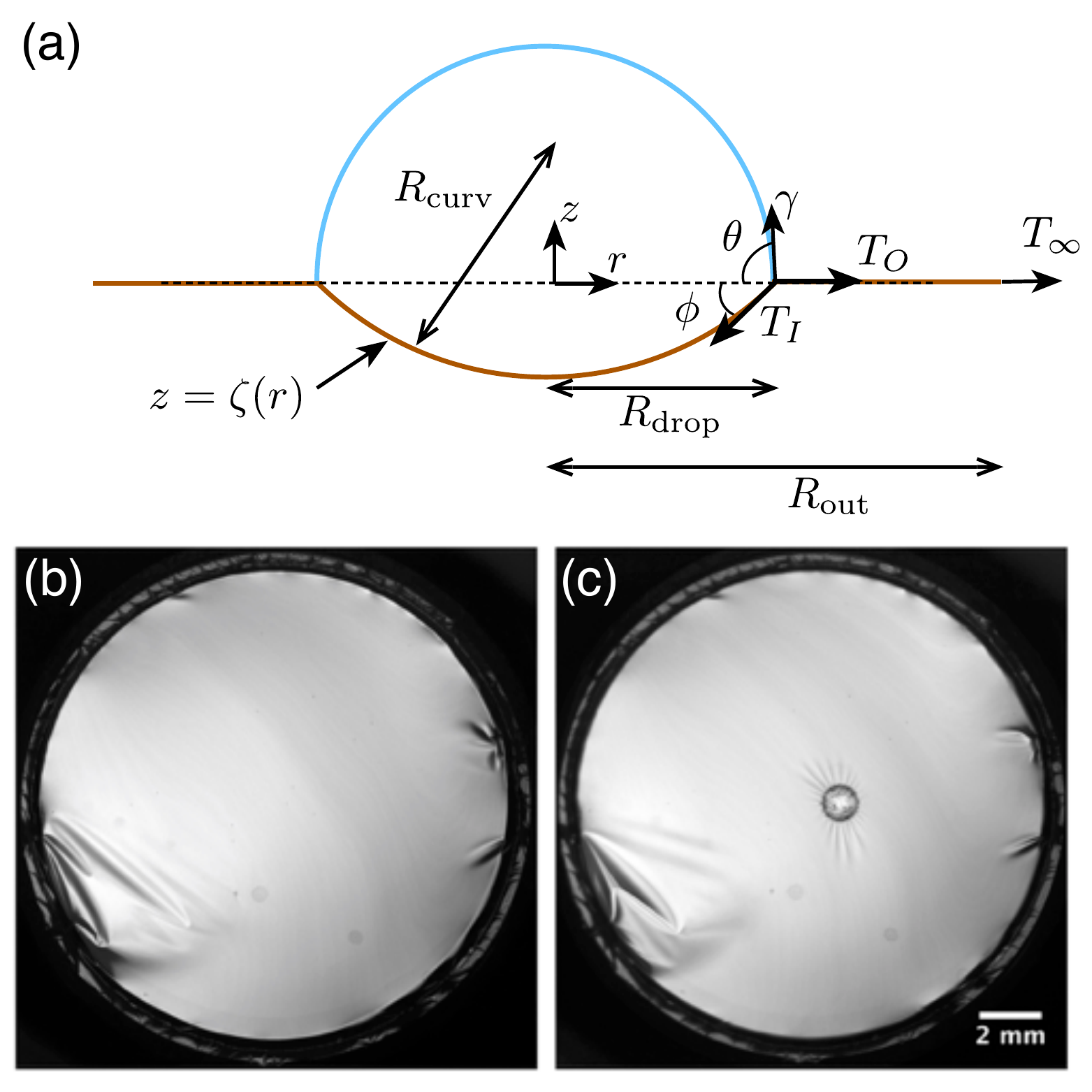}
\caption{ A droplet placed on a thin elastic membrane that is clamped at its boundary, $r=\Rout$. (a) Schematic showing the various dimensions of the system, together with an illustration of the forces that act at the contact line of the droplet. (b) Plan view of a Polystyrene membrane (thickness $t=364 \mathrm{~nm}$) clamped at its edges on the wall of a cuvette. (c) After a small drop of water is deposited on the surface of the PS membrane, wrinkling is observed around the periphery of the droplet. The droplet's volume is sufficiently small ($V/\ell_c^3\ll1$, where $\ell_c = (\glv/\rho g)^{1/2}$ is the capillary length) and so the formation of wrinkles cannot be attributed to the normal force exerted on the sheet by the drop's weight \cite{Huang07,Vella15}; instead wrinkling indicates a capillary-induced tensile force that is sufficiently strong to pull the dry part of the sheet towards the contact line. The radial wrinkles relax the consequent hoop compression. (Images in (b) and (c) courtesy of Deepak Kumar, UMass Amherst.)}
\label{fig:setup}
\end{figure}

In this paper, we introduce a comprehensive theory of this problem, building on and expanding previous work by Schroll \emph{et al.}~\cite{Schroll13}.  
A central result of our analysis is that the effect of the liquid drop is generally 
{\emph{non-perturbative}}: the presence of the drop is {\emph{not}} a small 
perturbation of the state of stress within the sheet prior to wetting 
\footnote{In this paper, we use the terminology of singular perturbation theory in which a {\emph{``non-perturbative effect"}} refers to a large, localized departure of a physical observable (here, the stress within the sheet) from its value in the ``unperturbed" state (here,  the uniformly-stretched, dry sheet).  Using this terminology emphasizes that adding an apparently small perturbation to the system  can  have a disproportionately large effect: here,  the drop surface energies, $\gamma_{sl},\gamma_{lv}$,  may be smaller than the stress within the dry sheet and yet the change in the stress due to the drop's presence may be large.}. 
To give a first indication of why this might be the case, we note that the capillary pressure of the drop  causes a parabolic bulge 
of the sheet. The presence of such a parabolic shape, with non-zero Gaussian curvature,  induces a tensile strain in the sheet, which can be approximated as $\epsilon\propto \phi^2$ (with $\phi$ the angle of the sheet at the interior of the contact line, see fig.~\ref{fig:setup}). The stress at the contact line may thus be approximated as $\Ti\sim Y\phi^2$. Vertical force balance at the contact line, however, implies  $\Ti\sim \gamma\sin\theta_Y/\phi$ (where we  assume for the moment that $\phi\ll1$ and that $\theta\approx\theta_Y$, so that the system is `close' to the YLD law). 
Equating these two estimates of $\Ti$, we find that
\begin{equation}
\Ti \sim (\sin\theta_Y \ \gamma)^{2/3} Y^{1/3} \ .
\label{eq:non-perturb}
\end{equation} Since the typical ratio $Y/\gamma\gg 1$, for many thin sheets including graphene \cite{Lee2008}, stiff polymers \cite{Schulman15}, and most elastomers \cite{Nadermann13},  the scaling (\ref{eq:non-perturb}) implies that $\gamma\ll\Ti\ll Y$. This tension may therefore dominate any pre-existing tension and the presence of the drop is, in general, 
\emph{not} a small perturbation to the pre-tension. It is also noteworthy that the effect of the drop is not related to its size, and thus occurs even for arbitrarily small drops (although the spatial extent of the effect \emph{is} proportional to the drop size). 

Our main concern in this paper is to place the theoretical understanding of recent experiments in this field on a firm footing. In particular, we shall use more detailed scaling arguments together with calculations from first principles to remove the reliance on any assumptions underlying the derivation of Eq.~(\ref{eq:non-perturb}) and thereby identify the parameter regimes in which the capillary effect is perturbative (with respect to pre-existing tension), or non-perturbative
(yielding the scaling \eqref{eq:non-perturb}). Furthermore, we will obtain quantitative expressions for the contact angles and stress in a partially-wet sheet as a function of its elastic moduli, interfacial tension, and any pre-tension associated with boundary loads or clamping at the sheet's far edge.

Our theoretical approach follows  a recent study that addressed this type of elasto-capillary mechanics for the partial wetting of a thin, circular solid sheet, floating on a liquid bath \cite{Schroll13}. In this approach, the mechanical equilibrium of the sheet is obtained by minimizing the total energy of the system, $\Usurf + \Ustrain + \Ubend$, and is described through the F\"{o}ppl-von-K\'{a}rm\'{a}n (FvK) equations, accounting for a tensile load, $\Tedge$, at its far edge (exerted by the surrounding liquid bath), and to capillary pressures and forces (exerted by the liquid drop). In that problem, solutions to the partial wetting problem were described by two primary dimensionless groups: 
\begin{gather}
{\rm (capillary)  \ bendability:} \  \epscap^{-1} \equiv \frac{\gamma \Rin^2}{B} \ 
\label{eq:define-bendability}, \\
{\rm (capillary) \ extensibility:} \ \tglv =\glv/Y  \ , 
\label{eq:DGSchroll}
\end{gather}
where $B \sim Et^3$ is the bending  modulus of the sheet. The bendability parameter can be understood as the ratio between the drop's radius, $\Rin$, and the elasto-capillary length, $\lbc=(B/\gamma)^{1/2}$ (called the ``bendo-capillary'' length by Style \emph{et al.}~\cite{Style17}); the extensibility parameter is the characteristic 
tensile strain induced in the sheet by capillary forces, as already discussed. The partial wetting of a highly-bendable yet nearly-inextensible sheet is thus described by a {\emph{singular}} 
limit ($\epscap \ll 1, \tglv \ll 1$), of the FvK equations. 

\begin{figure}
\centering
\includegraphics[width=0.9\columnwidth]{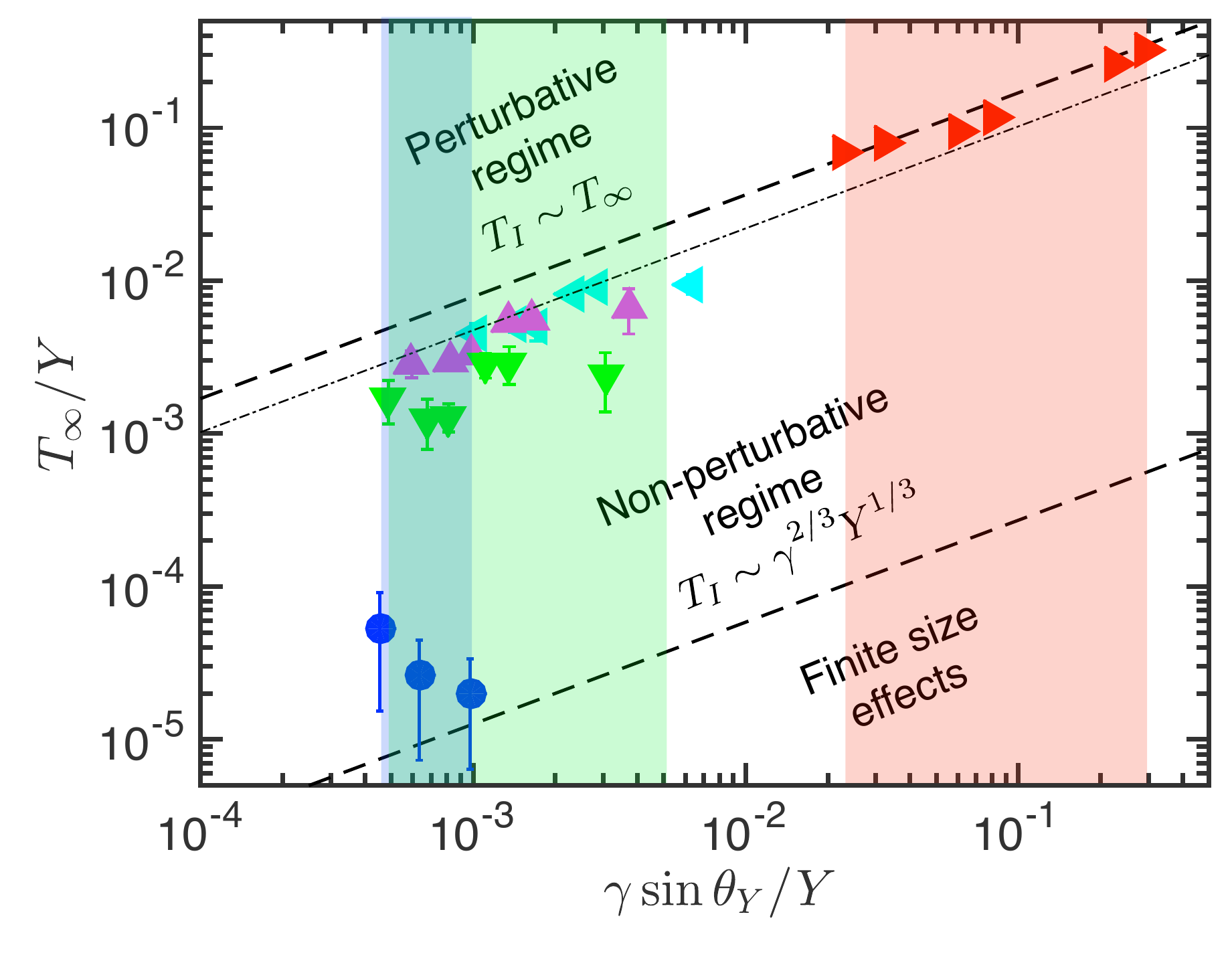}
\caption{Diagram showing the regions of parameter space in which the presence of a drop has a perturbative or non-perturbative effect on the stress state within the sheet (defined as a change in the tension at $r=\Rin$ of less than or more than $10\%$, respectively, from the uniform, isotropic tension in the sheet prior to placing the drop, and shown by the upper dashed line; the dash-dotted line shows the same concept but for a $40\%$ change). The rectangles show the parameter space that has been explored experimentally for glycerol drops on polymeric sheets (SIS, $E \approx 0.8~\mathrm{MPa}$) \cite{Schulman15}, glycerol drops on glassy sheets (PnBMA, $E \approx 1~\mathrm{GPa}$) \cite{Schulman15} or various drops on polymeric sheets (PDMS, for which we take $E \approx 2.85~\mathrm{MPa}$ here) \cite{Nadermann13}. Our estimation of the associated far-field tension $\Tinf$ in these experiments (using a methodology we describe in \S4) is given by points as follows: Glycerol drops  \cite{Schulman15} on  PnBMA films (blue circles) and SIS films (red right pointing triangles); experiments with PDMS films \cite{Nadermann13} and drops of De-ionized water (cyan left pointing triangles), Ethylene-Glycol (magenta upward pointing triangles) and DMSO (green downward pointing triangles).  Here the region where finite size effects become important has been calculated with $\Rout/\Rin=100$.}
\label{fig:RegimeDiagSlivers}
\end{figure}

In parallel to the experimental and theoretical studies of partial wetting on floating sheets \cite{Huang07,Schroll13,Toga13}, several groups have addressed the partial wetting problem in a different, but related, setup:  a sheet is suspended in vapour while subject to some unknown pre-tension, $\Tpre$, and clamped at its far edge.  A small liquid droplet is then placed on the (clamped) sheet \cite{Nadermann13,Schulman15}.  Although the characteristic values of the bendability and extensibility parameters~\eqref{eq:DGSchroll} in these experiments are similar to those in  studies of floating sheets  \cite{Huang07,Schroll13,Toga13}, the possibility that 
the capillary-induced stress might exhibit a non-trivial scaling, such as that in \eqref{eq:non-perturb}, appears to have been overlooked \cite{Nadermann13,Schulman15}.
The potential impact of this neglect of the non-perturbative nature of the problem is illustrated in fig.~\ref{fig:RegimeDiagSlivers}, which shows the parameter regimes (in terms of the normalized surface tension $\gamma/Y$ and  tension far from the drop, $\Tinf/Y$) for which the tension at the contact line is a small perturbation of the far-field tension $\Tinf$ ($\Ti\approx\Tinf$ to within an error of  $10\%$) or rather has a non-perturbative effect. (For sufficiently large sheets we expect that $\Tpre\approx\Tinf$.) This figure also shows the values of the parameter $\tglv=\glv/Y$  investigated experimentally \cite{Nadermann13,Schulman15} as filled, coloured columns. From this figure we see that 
partial wetting of 
both the PnBMA sheets used by Schulman \& Dalnoki-Veress~ \cite{Schulman15} and the PDMS sheets used by Nadermann \emph{et al.}~\cite{Nadermann13} may be described by a perturbative theory only if the pre-tension is at least an order of magnitude larger than $\glv$. However, using data from Refs.~\cite{Schulman15,Nadermann13} combined with our detailed FvK-based calculations (see \S 4) suggests that the actual far-field tensions, $\Tinf$, (shown as green, cyan, magenta, and blue symbols) may not reach such large values. This places these experiments in the non-perturbative regime, requiring one to account for the effect of a capillary-induced tension, Eq.~(\ref{eq:non-perturb}).

\begin{figure}
\centering
\includegraphics[width=0.85\columnwidth]{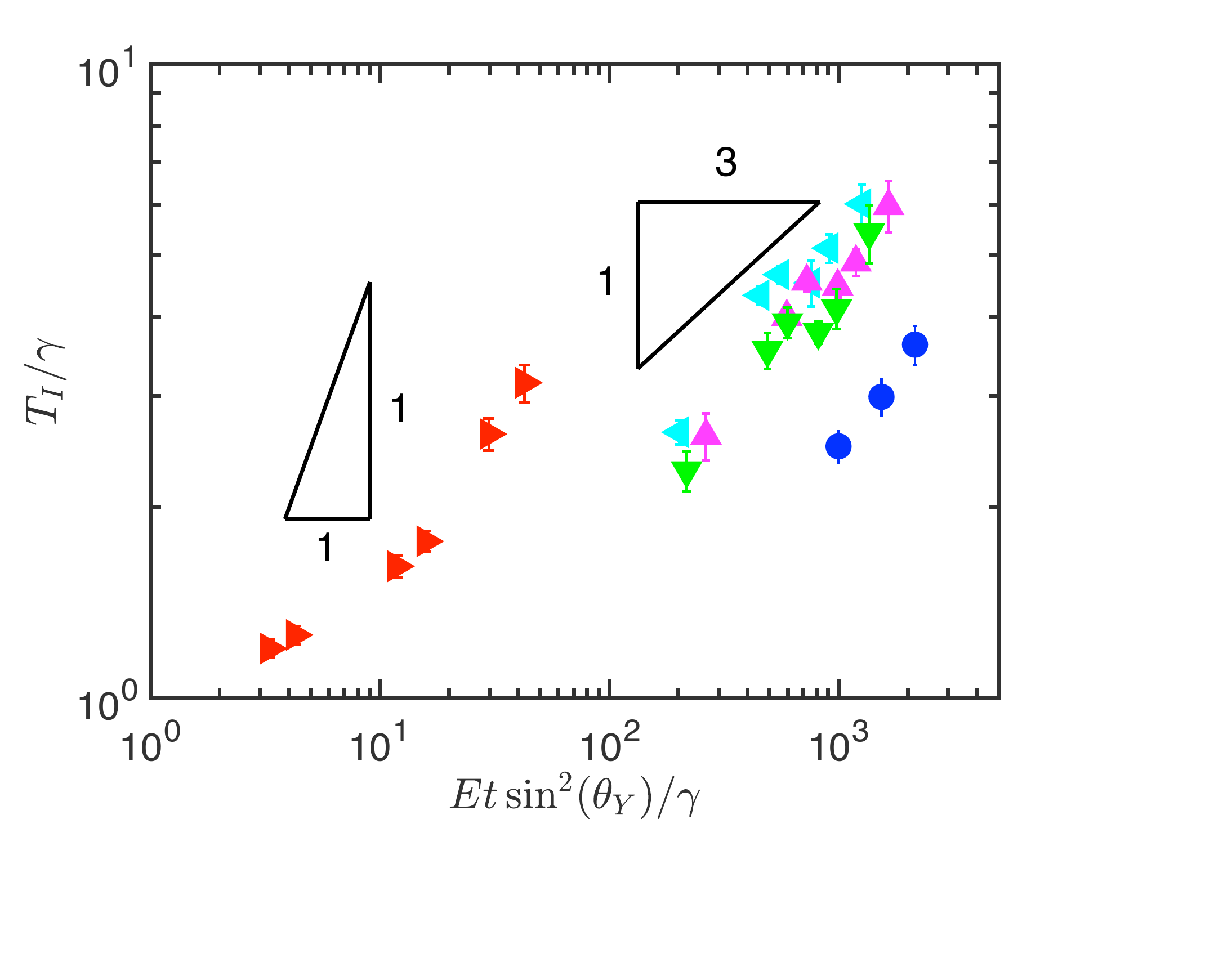}
\caption{The dependence of the measured internal tension, $\Ti \approx \glv \sin\theta_Y/\phi$, on the sheet's thickness $t$. Here the results of experiments performed by Schulman \& Dalnoki-Veress \cite{Schulman15} are combined with the results of experiments of Nadermann \emph{et al.}~\cite{Nadermann13}. Data are shown for Glycerol droplets \cite{Schulman15} on  PnBMA films (blue circles) and SIS films (red right pointing triangles); other triangles show  results  for PDMS sheets  \cite{Nadermann13} with different liquid droplets as follows: De-ionized water (cyan left pointing triangles), Ethylene-Glycol (magenta upward pointing triangles) and DMSO (green downward pointing triangles). (Throughout we use the same Values of Young's modulus, $E$, as in Fig.~2.) We note that results for PnBMA and PDMS sheets appear to be consistent, over the range of values plotted, with the
scaling $\Ti\sim t^{1/3}$, expected for a material with a thickness-independent Young modulus $E$ in the non-perturbative regime --- in this regime $\Ti$ is dominated by the capillary-induced stress, $T^\ast \sim \glv^{2/3} Y^{1/3}$, which is much larger than the pre-tension in the sheet. Furthermore, for the PDMS data, the dependence of $\Ti$ on the YLD angle $\theta_Y$ (which takes different values for the various types of liquid used in the experiments  \cite{Nadermann13}) appears to be consistent with the scaling relation (\ref{eq:non-perturb}) predicted in the non-perturbative regime. (Our quantitative solution of the FvK equations (\S4) shows that in the non-perturbative regime the actual value of $\Ti$ exhibits a  
very weak dependence, $\Ti \sim 1/\log(T^{\ast}/\Tpre)$, of $\Ti$ on the pre-tension, $\Tpre$.
Note that our analysis is unable to determine the {\emph{source}} of the pre-tension in the sheet, and thus cannot explain the noticeable vertical shift between data sets that correspond to PnBMA  \cite{Schulman15} and PDMS sheets  \cite{Nadermann13}. Since the pre-tension is generated upon clamping the dry sheet (prior to placing the drop), it may be governed by nontrivial solid chemistry and the preparation techniques used (see also the discussion in \S5.2).
Note also that the $x$-axis here is the inverse of the capillary strain, and so is required to be large for our analysis to be valid --- this may explain why the thinnest SIS sheets deviate from the linear scaling $\Ti\sim t$, expected in the perturbative regime ($\Ti \approx \Tpre$) for a thickness-independent  in-plane pre-strain.} 
\label{fig:TiRaw}
\end{figure}

The potential importance of the elasto-capillary tension scale $\glv^{2/3}Y^{1/3}$, and hence the non-perturbative character of the problem, is further emphasized in fig.~\ref{fig:TiRaw}. There we plot the value of the tension in the vicinity of the contact line, $\Ti$, as a function of the  sheet  thickness (using experimental data from Nadermann \emph{et al.}~\cite{Nadermann13} and Schulman \& Dalnoki-Veress \cite{Schulman15}). This replotting of 
data also suggests that experiments with PDMS \cite{Nadermann13} and PnBMA \cite{Schulman15} films may lie in the non-perturbative regime: the observed tension
appears to be consistent with the non-perturbative scaling $\Ti\sim (\sin\theta_Y\ \glv)^{2/3}Y^{1/3}$ (Eq.~\ref{eq:non-perturb}), suggesting that the role of any pre-existing tension may be small. (Experiments with SIS films \cite{Schulman15} do not exhibit this scaling and, hence, are likely to lie in the perturbative regime.) We shall therefore argue that proper consideration of the non-perturbative effect of the capillary-induced stress, $T^\ast=\gamma^{2/3}Y^{1/3}$, should be given when interpreting the results of experiments in such scenarios. In particular, rather than invoking a constitutive relation for $\Ti(t)$ (such as the linear relation assumed in Ref.~\cite{Nadermann13}),    
our FvK-based theory explains the measured values of $\Ti$ in Ref.~\cite{Nadermann13} as the non-perturbative, capillary-induced contribution to the stress in the partially-wet sheet, $\Ti\sim T^\ast$. 

In this manuscript we seek to bridge the studies of Refs~\cite{Huang07,Schroll13,Toga13} and Refs~\cite{Nadermann13,Schulman15}, analyzing the partial wetting of highly bendable, nearly inextensible solid sheets in a general way, 
applicable to sheets that are either clamped with pre-tension $\Tpre$, or are subject to a fixed tensile load $\Tedge$ at their far edge. Correspondingly, we extend the theoretical analysis of Schroll \emph{et al.}~\cite{Schroll13} in three key directions. First, we provide a qualitative, scaling-type analysis that elucidates the origin of the non-perturbative capillary-induced stress (\ref{eq:non-perturb}) and demonstrates its relevance to 
any 
far-field boundary conditions (BCs) 
in a  
particular experimental setting. 
Second, we introduce the dry tension
\begin{equation}
\Tdry =  
\left\{ \begin{array}{ll} \Tedge & ({\rm load \ controlled}) \\ \\ \Tpre & ({\rm clamping})  \end{array} \ , \right. 
\label{eq:defineTdry}
\end{equation} which in turn leads to the definition of a `dry extensibility':
\begin{equation}
\tTdry =  \Tdry/Y\ .
\label{eq:define-dry-exten}
\end{equation}
To simplify the analysis,
 we introduce an `effective far-field tension', $\Tinf$. In the case of large sheets, $\Rout/\Rin\gg1$, we shall see that $\Tinf\approx\Tdry$.  (The key observation for now, however, is that the ratio $\glv/\Tinf$ may take arbitrary values, in contrast to the analysis of Schroll \emph{et al.}~\cite{Schroll13}, which considered only  $\glv/\Tinf = O(1)$.)
The third, and final, difference with the work of Schroll \emph{et al.}~\cite{Schroll13} is that we show that the dominant behaviour of the capillary-induced stress, Eq.~(\ref{eq:non-perturb}), 
and contact angles, $\phi$ and $\theta$, as $\glv/Y\to0$, may be predicted without resorting to
an explicit  minimization of the total energy,  simplifying the computational method considerably.  
Our generalized analysis allows us to identify the parameter regimes for which the capillary-induced stress is either perturbative or non-perturbative, and thereby use our theoretical results to revisit the assumptions made in recent experimental works and to reassess the conclusions drawn from their experimental data.

\subsection{Outline}

We commence, in \S2, with a qualitative discussion of the subtle interaction between pre-existing tension, the inherent resistance of a solid sheet to stretching and the presence of a liquid drop. Using energetic considerations and scaling arguments, we explain our central conclusions concerning the drop-induced tension without using the FvK equations.  In \S3 we discuss the full FvK model of the problem, focusing on the BCs that are appropriate at the far edge and in the vicinity of the contact line \footnote{We note that Nadermann \emph{et al.}~\cite{Nadermann13} also employed an FvK-based analysis, which led them to conclude that the stress in the partially-wet sheet is governed by a pre-tension, as long as the sheet is ``sufficiently thin''. We address the disagreement between our theoretical approach and that of Nadermann \emph{et al.}~\cite{Nadermann13} in Appendix C.}. The axial symmetry of the geometry leads us to seek an axially symmetric solution of the FvK equations. Our solution of this problem reveals that in some regimes radial wrinkles may form, though the system may still be described in an axisymmetric setting using tension field theory as the asymptotic limit of the FvK equations in the singular, high bendability limit ($\epscap \ll 1$) \cite{Davidovitch11,King12,Schroll13}. In \S4 we discuss the solution of the FvK equations, focusing on large sheets, and small values of the extensibility parameters; our results are shown in figs~4-8. In \S5 we discuss 
some recent works \cite{Vella10,Nadermann13,Schulman15} that addressed the partial wetting of solid sheets under tension, and the implications of our findings for the interpretation of experimental data. We also highlight some unresolved questions.

{\emph{A reader who is not familiar with the FvK theory, may skip \S3 at a first reading, focusing instead on the scaling analysis in \S2 and  the discussion and critique presented in \S4 and \S5.}} \\

\underline{\emph{Nomenclature:}} 

$\bullet$ 
A primary object of our study is the deviation of the angles $\theta,\phi$, together with the stress discontinuity, $\To -\Ti$ (Fig.~\ref{fig:setup}), from the values assumed in the YLD law (\ref{eq:LaplaceLaw}). As will be revealed in the sequel, such an analysis requires us to distinguish between the ``geometry" ({\emph{i.e.}}~the contact angles), and the ``mechanics" ({\emph{i.e.}}~the stress discontinuity) underlying the YLD law (\ref{eq:LaplaceLaw}). Hence, we will use the terms  ``YLD contact geometry" and ``YLD contact mechanics" as handles for the following limits:   
\begin{subequations}
\label{eq:YLDcontact-define}
\begin{equation}
{\rm{  YLD  \ contact  \ geometry:}} \ \ \theta \to \theta_Y  \ \  ; \ \   \phi \to 0 \  , 
\label{eq:YLDgeometry-define}
\end{equation} 
and: 
\begin{equation}
{\rm{  YLD  \ contact  \ mechanics:}} \ \ \To -\Ti \   \to  \ \glv\cos\theta_Y   \ , 
\label{eq:YLDmechanics-define}
\end{equation}
\end{subequations} where the symbol ``$\to$" refers to (various orders of) the asymptotic limit, $\tglv, \tTdry \to 0$, of vanishing extensibility parameters, defined in Eqs.~\eqref{eq:DGSchroll} and \eqref{eq:define-dry-exten}, respectively.

$\bullet$ 
Throughout our analysis, we shall need to consider various tensions within the elastic sheet, as well as interfacial tensions. We shall adopt the convention that tensions within the sheet are denoted $T_{(\cdot)}$, while interfacial tensions are denoted $\gamma_{(\cdot)}$. Similarly, in some circumstances it is the size of a tension compared to the stretching modulus, $Y=Et$, that is important, while in other circumstances it is its size relative to the capillary-induced tension, $T^\ast=\glv^{2/3}Y^{1/3}$, that is more pertinent; we shall therefore let $\tilde{(\cdot)}=(\cdot)/Y$ while $\bar{(\cdot)}=(\cdot)/T^\ast$.

\section{Qualitative discussion\label{sec:qualitative}}

\subsection{Scaling analysis}

To further elucidate the non-perturbative nature of a liquid drop sitting on a solid sheet, we commence with a qualitative scaling analysis, based on energetic considerations.
For this purpose, we assume that the dry portion of the sheet, $\Rin<r<\Rout$ with $\Rout \gg \Rin$, remains flat,  while its wet part, $0\leq r<\Rin$, approximates a spherical cap with some radius of curvature $\Rcurv$. We further assume that the stress field in the wet portion of the sheet ($r \lesssim \Rin$) is characterized by some tensile scale $\Twet$, which must be larger than $\Tdry$, and that in the dry portion of the sheet the stress returns to the isotropic, spatially uniform value $\Tdry$. As we will show, defining: 
\begin{equation}
\dTcap \equiv \Twet-\Tdry \ , 
\label{eqn:dTcap}
\end{equation} this simplified picture suffices to characterize the parameter regime at which 
the scaling law (\ref{eq:non-perturb}) is valid. 

\emph{Base state:} Before introducing the drop, the dry sheet is subject to a uniform tension $\Tdry$. The elastic energy of this dry state is then  $\pi\Rout^2(1-\nu) \Tdry^2/Y$ with $\nu$ the Poisson ratio of the sheet; the combined surface energy of the dry sheet and the liquid drop is $2\gamma_{sv} \pi \Rout^2 + 4\pi \glv \Rin^2$.\\

We now evaluate the energetic benefit of the partially wet state in comparison to this non-contacting ``base" energy.            
\\

 {\emph{Surface energy of the wet state:}} The reduction in surface energy, $\Delta \Usurf$, is determined by the contact line radius, $\Rin$, the surface tension $\glv$, and by the difference between the surface energies of the solid surface when wet or dry, $\Delta\gamma_s = \gamma_{sv} - \gamma_{sl}$. Hence, the gain in surface energy can be written
\begin{equation}
\Delta \Usurf  = \glv \Rin^2 G(\phi,\theta;\theta_Y)  \ , 
\label{eq:USurf}
\end{equation}
where $G(\phi,\theta; \theta_Y)$ is a function of the two unknown angles, $\phi, \theta$ (fig.~1b), and the dimensionless parameter $\theta_Y = \cos^{-1}(\Delta \gamma_s/\glv)$. For our work, the key feature of the function $G$ is that when the solid surface is  perfectly flat (i.e.~$\phi=0$), then  $G(\phi=0,\theta;\theta_Y)$ is known to be minimized by $\theta=\theta_Y$, to recover the YLD law.
\\

{\emph{Elastic energy of the wet state:}} While wetting reduces the surface energy of the system, it introduces a Gaussian curvature, and therefore increases the elastic energy of the sheet. This energetic cost, $\Delta \Ustrain$, may be written in scaling terms as:
\begin{equation}
\Delta \Ustrain \sim \Rin^2 \frac{\Twet^2 -\Tdry^2}{Y}  \  +  \ Y \Rin^2 \left(\frac{\Rin}{\Rcurv}\right)^4   
\label{eq:StrainScaling-1}
\end{equation}

To understand the difference between the two terms on the RHS of \eqref{eq:StrainScaling-1} (and why considering their sum does not amount to ``double counting" a single cost), consider the problem of placing a disk-like sheet of radius $\Rin$ onto a rigid ball of radius $\Rcurv$ by pulling the perimeter of the sheet with a radial tension $\Twet$ \cite{Hure11,Hohlfeld15}. The strain can be written schematically as $strain = (\Twet/Y) + (\Rin/\Rcurv)^2$, where the first term is a ``mechanical" contribution (which would exist even if the sheet  remained planar), and the second term is a ``geometric" contribution (which would exist even if the sheet were forced to wrap the ball, without pulling its edge). The RHS of Eq.~\eqref{eq:StrainScaling-1} is then recognized as the difference between the energy $\sim Y (strain)^2 \Rin^2 $ and the strain energy of the disk in the base (dry) state $\sim Y (\Tdry/Y)^2 \Rin^2$.

Finally, the direct cost of bending energy, $\Ubend$, that the drop imposes on the sheet, can be estimated as:
\begin{equation}
\Ubend\sim B\Rin^2/\Rcurv^2  \ , 
\label{eq:Ubend-00}
\end{equation}
which is much smaller than the energetic cost of straining the sheet, and can thus be safely ignored.  
More precisely, the ratio $\Ubend/\Ustrain$ 
scales as a (negative) power of the bendability parameter $\epsilon^{-1}$ defined in (\ref{eq:define-bendability}), and is thus negligible in our high-bendability analysis. (More details on the scaling of $\Ubend/\Ustrain$ with $\epsilon^{-1}$ may be found in Appendix B.)

Recalling that we are assuming that the characteristic tensile strains are small ({\emph{i.e.}} $\tglv, \tTdry \ll 1$), we can compare the geometric contribution to $\Delta \Ustrain$ with $\Delta \Usurf$;  one immediately sees that for the energetic cost (due to strain) to not exceed the gain in surface energy, requires $(\Rin/\Rcurv)^2 \to 0$ as $\tglv \to 0$. Simple trigonometry then shows further that, in this limit, $\sin\phi \approx \Rin/\Rcurv$.

Our scaling analysis therefore reveals that the angle of the sheet at the meniscus $\phi \to 0$ in the nearly inextensible limit, $\tglv \to 0$. Finally, anticipating (and confirming below in a self-consistent manner), that in such a contact geometry the decrease in surface energy substantially exceeds  the energetic cost of strain (namely, $\Delta \Usurf /\Delta \Ustrain \to \infty$ as $\tglv \to 0$), we are left with the problem of minimizing the energy $\Delta \Usurf$, Eq.~\eqref{eq:USurf}, subject to the condition that $\phi \to 0$. We therefore conclude that:
\\

{\emph{In a partially wet solid sheet,  the YLD contact geometry (\ref{eq:YLDgeometry-define}) 
is approached asymptotically in the nearly inextensible limit:}} 
$$\phi \to 0, \theta \to \theta_Y   \ \ \text{as}  \ \ \tglv \to 0  \ . $$

The above result does not suffice to determine the exact ``rate" at which the YLD contact geometry is approached (namely, how $\phi$ and $\theta-\theta_Y$ scale with $\tglv$). To determine this, and thereby the tension  in the wet portion of the sheet, $\Twet$, we note that vertical force balance at the contact line implies: 
\begin{equation}
\Twet = \glv \frac{\sin\theta}{\sin\phi} \ . 
\label{eq:scaling-normal}
\end{equation}
Hence, the smaller $\phi$ is, the larger is the tensile load induced in the sheet and the energetic cost associated with the mechanical strain, $\Twet/Y$. If the  
mechanical and geometric sources of strain (the two terms on the RHS of Eq.~\ref{eq:StrainScaling-1}), were to balance each other, then one obtains, with the aid of the geometric relation $\phi\sim \Rin/\Rcurv$, that $\phi\sim(\glv/Y)^{1/3}\ll 1$, 
and 
$\Twet\sim T^\ast$ where: 
\begin{equation}
T^\ast(\glv,Y) \sim Y^{1/3} \glv^{2/3}.
\label{eq:T-transit}
\end{equation}
The estimate, $\Twet \sim T^\ast$ ignores the effect of $\Tdry$; the size of $\Tdry$ in comparison to $T^\ast$  therefore determines two, qualitatively different, regimes: 
\\

{\underline{\bf Non-perturbative regime ($T^\ast \gg \Tdry  $):}} 

\noindent
If the tension  within the dry portion of the sheet (away from the contact line), $\Tdry$, is sufficiently small in comparison to $T^\ast$,  
the first term on the RHS of Eq.~(\ref{eq:StrainScaling-1}) may be safely approximated as $\Rin^2 {\Twet^2}/{Y}$. The balance just discussed then yields: 
\begin{equation} 
\dTcap  = \Twet - \Tdry \approx \Twet \sim \glv^{2/3}Y^{1/3} \sim T^\ast \gg \Tdry
\label{eq:nonperturb-100}
\end{equation} 
such that the YLD law is recovered asymptotically: 
\begin{equation}
\phi \sim \tglv^{1/3} = (\glv/Y)^{1/3} \ \ ; \ \ \theta \to \theta_Y.
\label{eq:approachYoung}
\end{equation}

{\underline{\bf Perturbative regime ($ T^\ast \ll \Tdry$):}} 

\noindent In contrast, if the tension within the dry part, $\Tdry$, is sufficiently large in comparison to $T^\ast$, the tension $\Twet$ in the wet part may be assumed to be only slightly larger than $\Tdry$. In such a case, the vertical force balance (\ref{eq:scaling-normal}) yields: 
\begin{equation} 
\phi \sim \glv/\Tdry  \ \ ; \ \ \theta \to \theta_Y. 
\label{eq:approachYoung-2}
\end{equation} 
In this case, the capillary-induced stress is small in comparison to the dry tension, and one obtains: 
\begin{equation} 
\dTcap  = \Twet - \Tdry \sim \glv \ll \Tdry. 
\label{eq:perturb}
\end{equation}
This demonstrates the perturbative nature of partial wetting if the tension in the dry part of the sheet is sufficiently large. 
\\

Inspecting the above results, a few comments are worth making. First, the parameter that determines whether the drop-induced stress, $\dTcap$, is large (non-perturbative) or small (perturbative) in comparison to the stress in the dry part, $\Tdry$, is the ratio $\Tdry/(Y^{1/3}\glv^{2/3})$. Thus, even if the drop is extremely small in comparison to the size of the sheet ($\Rin/\Rout \ll 1$), the drop's effect on the stress in the sheet in the vicinity of the contact line may be very large. Second, the results of the scaling analysis are insensitive to the method by which the tension $\Tdry$ is established away from the drop (i.e.~clamping or fixed load). Third, the YLD contact geometry (\ref{eq:YLDgeometry-define}) may be approached even if the tension $\Tdry$ in the dry part is very small in comparison to the surface tension $\glv$, provided that $\glv/Y\ll1$. 
Note that this is another mechanism through which the YLD contact geometry may be reached in thin sheets, in addition to the limit $\Tdry/\gamma\to\infty$, which was pointed out by Schulman \& Dalnoki-Veress \cite{Schulman15}. Finally, as the sheet approaches the inextensible limit (decreasing values of 
$\tglv, \tTdry$), the minimal relative tension required for the effect of the drop to be perturbative  diverges, namely, $T^\ast/\glv \sim \tglv^{-1/3} \to \infty$.

\subsection{Energetic hierarchy} 

Our scaling analysis relied on the assumption that as $\tglv \to 0$, the ratio $\Delta \Usurf/\Delta \Ustrain \to \infty$. The validity of this assumption can  now be verified self-consistently by substituting  the scaling results for $\Twet$ and $\phi \approx \Rin/\Rcurv$ into the RHS of Eq.~\eqref{eq:StrainScaling-1}. 
Using Eqs.~(\ref{eq:non-perturb},\ref{eq:approachYoung}) for the non-perturbative regime ($\Tdry<T^\ast$), and Eqs.~(\ref{eq:approachYoung-2},\ref{eq:perturb}) for the perturbative regime ($\Tdry>T^\ast$), we find  $\Delta \Ustrain \sim \Rin^2 (\glv^4/Y)^{1/3}$ and $\Delta \Ustrain \sim \Rin^2 [\Tdry \glv/Y \ + \  Y (\glv/\Tdry)^4]$, respectively. The relevant asymptotic ratio in the non-perturbative regime is then
\begin{equation}
{\rm For} \  \Tdry<T^\ast: \ \ \frac{\Delta \Usurf}{\Delta \Ustrain} \sim \left(\frac{Y}{\glv}\right)^{1/3} \ ,  
\label{eq:Uratio-non-perturb}
\end{equation} 
whereas in the perturbative regime: 
\begin{equation}
\frac{\Delta \Usurf}{\Delta \Ustrain} \sim \begin{cases}
 \frac{\Tdry^4}{Y\glv^3},\quad T^\ast<\Tdry < T^\ast\left(\frac{Y}{\glv}\right)^{1/15},\nonumber\\
 \frac{Y}{\Tdry},\quad \Tdry> T^\ast\left(\frac{Y}{\glv}\right)^{1/15}.
\end{cases}
\label{eq:Uratio-perturb}
\end{equation} 
\\ 

Notwithstanding these various asymptotic limits, the key point is that in each case the partial wetting of a highly-bendable, nearly-inextensible sheet
exhibits an energetic hierarchy: 
\begin{equation}
\Delta \Usurf  \ \gg \ \Delta \Ustrain \gg \Ubend \ . 
\label{eq:UHeirarchy}
\end{equation}
Similarly to the classical YLD law, 
for which the only energy of relevance  is $\Usurf$, the energetic hierarchy of a partially wet thin sheet under tension is characterized by dominance of the surface energy, which underlies a geometric constraint (the asymptotically flat state of the sheet). However, the various asymptotic regimes are spanned by two scales of residual tensile strain: $\tglv$ and $\tTdry$; the ratio between these two small, but crucially different, scales give rise to distinct, nontrivial routes through which the YLD contact of (\ref{eq:YLDcontact-define}) is attained.   

Energetic hierarchies similar to (\ref{eq:UHeirarchy}) emerge in a host of problems in which a {\emph{Gaussian curvature}} is imposed on a thin solid sheet with the aid of tensile loads that pull on its edges. In such problems, a separation of energy scales of the type signified by Eq.~(\ref{eq:UHeirarchy}) is interpreted as an {\emph{``asymptotically isometric"}} response, whereby the elasticity of the sheet only enters through various types of asymptotic constraints on the value of a dominant energy (here the surface energy) that does not depend explicitly on any of the elastic parameters of the sheet. Other examples of the intricate mechanics associated with asymptotically isometric response include the indentation of floating sheets \cite{Vella15,Paulsen16} and pressurized shells \cite{Vella15EPL,Taffetani17}, the wrapping of liquid volumes with ``solid surfactants'' \cite{Paulsen15}, and the twisting of a pre-stretched ribbon \cite{Chopin13,Chopin15,Dinh16}. 

One must thus appreciate the crucial, intimate relation between the non-perturbative nature of the capillary-induced stress $\dTcap$ and the Gaussian curvature imposed by the drop on the wet part of the sheet. By ignoring  or underestimating   this geometric source of strain ({\emph{i.e.}} the second term on the RHS of Eq.~\ref{eq:StrainScaling-1}), one is misled to conclude that capillary-induced stress is necessarily a small perturbation to a pre-existing tension ($\Tdry$) in the sheet. We emphasize that this is a feature of the two-dimensional nature of this problem --- in the one-dimensional, but otherwise identical, versions of this phenomenon \cite{Py07,Neukirch13,Hui15}, the Gaussian curvature is identically zero, and this complexity does not exist.

\section{Quantitative theoretical approach\label{sec:details}}

\subsection{The F\"{o}ppl-von K\'{a}rm\'{a}n equations}

Our starting point for a quantitative analysis is the F\"{o}ppl-von K\'{a}rm\'{a}n (FvK) equations, which describe the mechanical equilibrium of solid sheets \cite{Landau66,Mansfield89}. The FvK equations are a nonlinear set of partial differential equations, but the axial symmetry of our problem 
allows their reduction to a coupled pair of ordinary differential equations (ODEs) for the 
%
vertical displacement of the membrane, $\zeta(r)$, and a stress potential, $\psi(r)$:
\begin{gather}
\frac{1}{r}\frac{\upd}{\upd r}\left(\psi\frac{\upd \zeta}{\upd r}\right)=p(r),
\label{eqn:FvK1} \\
r\frac{\upd}{\upd r}\left\{\frac{1}{r}\frac{\upd(r\psi)}{\upd r}\right\}=-\frac{Y}{2}\left(\frac{\upd\zeta}{\upd r}\right)^2 \ .
\label{eqn:FvK2}
\end{gather} 
Here  the loading $p(r)$ is due to the capillary pressure within the droplet (for $0\leq r\leq\Rin$) and vanishes for $\Rin<r\leq\Rout$. The radial and hoop stresses in the sheet are related to the stress potential $\psi(r)$ through $\srr=\psi(r)/r$, and $\sqq=\upd \psi/\upd r$, which ensures that the in-plane equilibrium of stresses holds automatically \footnote{Note that the FvK equations (\ref{eqn:FvK1},\ref{eqn:FvK2}) determine the actual, thickness-integrated stress (force/length) within the sheet: $\sigma_{ij} (r)/t$ 
is the average force/area in the direction $\vec{e}_i$ ($i\in\{r,\theta\}$), on the face of an infinitesimal solid volume element within the sheet whose (outward) normal is in the direction $\vec{e}_j$ ($j\in\{r,\theta\}$). 
Hence, we need not consider the distinction (which appears often in the literature) between  ``elastic" and ``surface" contributions to the stress within the sheet \cite{Schroll13}. Note, however, that Eqs.~(\ref{eqn:FvK1},\ref{eqn:FvK2}) are the Euler-Lagrange equations derived from the elastic energy alone (which penalizes for bending and straining the sheet with respect to its rest, planar, stress-free state, but does not account for the energetic cost of surface area). Hence, since we pointed out already in \S1.1 that the energetically-favorable state of the system is obtained by minimizing the sum of energies, $\Ustrain + \Ubend + \Usurf$ \cite{Olives93}, the difference in solid surface energies, $\gamma_{sv} - \gamma_{sl} = \glv \cos\theta_Y$, does ultimately affect the stress within the sheet, even though it does not appear explicitly in Eqs.~(\ref{eqn:FvK1},\ref{eqn:FvK2}). As we explain in Appendix A (A.1.3), the term $\glv \cos\theta_Y$ enters through a boundary condition that is required to solve the set of ODEs (\ref{eqn:FvK1},\ref{eqn:FvK2}).}.
Equation (\ref{eqn:FvK1}) results from vertical  force balance on the sheet, in which we neglect the effects of the sheet's bending stiffness --- our analysis therefore corresponds to  ``membrane  theory" \cite{Mansfield89}, or (see Appendix A) to ``tension field theory" when the capillary effect is strong enough to induce (hoop) compression in the radially stretched sheet  \cite{Davidovitch11,Schroll13}.
The second equation, (\ref{eqn:FvK2}), expresses the compatibility of strains.

Turning to the $1^{st}$ FvK equation~(\ref{eqn:FvK1}), we note that 
Laplace's law relates the pressure in the droplet to the curvature of the liquid-vapor interface (see fig.~1), and hence:  
\beq p(r)=\begin{cases}
2\glv\sin\theta/\Rin,\quad 0\leq r\leq\Rin\\
0,\quad \Rin \leq r\leq \Rout.\end{cases}\eeq  \ 
Integrating Eq.~(\ref{eqn:FvK1})  once, we obtain:
\beq\psi\frac{\upd \zeta}{\upd r}=\begin{cases}\frac{\glv\sin\theta}{\Rin}r^2,\quad 0\leq r\leq\Rin\\
0,\quad \Rin< r\leq\Rout
\end{cases}
\label{eqn:FvK1Int}
\eeq 
where the constants of integration in both parts of Eq.~(\ref{eqn:FvK1Int}) vanish: the first to ensure that the membrane is flat at the centre, $\zeta'(0)=0$, and the second since there is no net vertical force on the sheet outside the drop, $r > \Rin$. \cite{Vella10}

\subsection{Eliminating membrane shape}

Equation~(\ref{eqn:FvK1Int})  
allows us to eliminate the membrane slope, $\zeta'$, from \eqref{eqn:FvK2}, to give  equations for the stress potential $\psi$ with no explicit dependence on the membrane shape $\zeta(r)$. 
Equation~\eqref{eqn:FvK2} becomes: 
\beq r\frac{\upd}{\upd r}\left\{\frac{1}{r}\frac{\upd(r\psi)}{\upd r}\right\}= -\frac{Y}{2}\left(\frac{\glv\sin\theta}{\Rin}\right)^2 r^4 \psi^{-2}, \quad 0\leq r\leq\Rin \label{eqn:FvK2-new-in} \eeq 
and 
\beq
\frac{\upd}{\upd r}\left\{
\frac{1}{r}\frac{\upd(r\psi)}{\upd r}  
\right\}=0,
\quad \Rin \leq r \leq\Rout\ . 
\label{eqn:FvK2-new-out}
\eeq 
Equation~(\ref{eqn:FvK2-new-out}) describes the Lam\'e problem \cite{Timoshenko70}: a solid annulus subject to a tensile load $\To$ at its inner edge, $\Rin$, and another load at its outer edge, $r=\Rout$; there are no stresses induced by out-of-plane displacements.
Writing the effective far-field tension as $\Tinf$, we find:
\beq
\psiout(r) =\Tinf r+ (\To - \Tinf)\frac{\Rin^2}{r}, \quad \Rin<r<\Rout \ .\label{eq:Lameaxi} 
\eeq 
For $\To<2\Tinf$, this classical Lam\'{e} solution yields purely tensile (i.e. positive) hoop and radial stresses; the sheet is therefore stable to out-of-plane deflections.
Note that at this stage the radial stress  in the sheet on the (dry) side of the contact line, $\To$, is not known; it must be  found by matching the solutions of the two equations (\ref{eqn:FvK2-new-in}) and (\ref{eqn:FvK2-new-out}), as we will describe in the sequel.

A key limitation of the Lam\'{e} solution \eqref{eq:Lameaxi} and the reduced FvK equation \eqref{eqn:FvK2-new-in} is that they are valid only if $\sqq>0$ everywhere. For a range of parameters, we find that $\sqq<0$  in an annulus $\Li<\Rin<\Lo$ around the contact line; wrinkles then emerge to relax this compressive stress. The presence of wrinkles changes the nature of the stress field qualitatively and requires special consideration, as discussed in Appendix A.

\subsection{Perturbative versus non-perturbative \label{sec:perturb}}

In Appendix A we discuss the steps necessary to obtain a complete solution to our problem. However, for the purposes of the forthcoming discussion, it is useful to assess the physical nature of the solution by re-writing the $2^{nd}$ FvK equation (\ref{eqn:FvK2-new-in}):
\begin{equation}
r\frac{\upd}{\upd r}\left\{\frac{1}{r}\frac{\upd[r\psi/(\Rin\Tinf)]}{\upd r}\right\}=-\tfrac{1}{2} \tauinf^{-3}\sin^2\theta\frac{r^4}{(\psi/\Rin\Tinf)^{2}}
\label{eqn:FvK2-new-in-n}
\end{equation} 
for $0\leq r\leq\Rin$, where the dimensionless parameter:
\beq
\tauinf=T_\infty/(\glv^2Y)^{1/3} \ 
\label{eq:tauinf}
\eeq compares the far-field tension, $T_\infty$,  to the capillary-induced tension generated by the combination of elasticity and surface tension, $T^\ast=\glv^{2/3}Y^{1/3}$, which we encountered already in \S2. Written in this way, one may readily distinguish between two distinct limits, demarcated by the value of the dimensionless parameter, $\tauinf$: 

{\emph{(i)}} If $\tauinf \gg 1$, the nonlinear term on the RHS of Eq.~(\ref{eqn:FvK2-new-in-n}), whose origin is the ``inhomogeneous source" term, $(\upd\zeta/\upd r)^2$ in Eq.~\eqref{eqn:FvK2}, is negligible. One  therefore expects $\psi$ to be well approximated by the homogeneous solution, 
$\psi \approx \Tinf r$. 
In this case, the capillary-induced stress has a perturbative effect on the in-plane stress in the sheet. This is as might be expected, since in this case the exerted tension, $\Tinf$, is significantly larger than that generated by the combination of surface tension and elasticity, $\glv^{2/3}Y^{1/3}$.

{\emph{(ii)}} In contrast, if   $\tauinf \ll 1$ 
the RHS of Eq.~(\ref{eqn:FvK2-new-in-n}) 
cannot be neglected. An elementary solvability condition is a  balance between the two sides of Eq.~(\ref{eqn:FvK2-new-in-n}), which yields the scaling, 
$\psi/\Rin \sim \gamma^{2/3}Y^{1/3}$. 
In this case, the tension, $\Tinf$, exerted on the sheet is significantly smaller than that generated by the combination of surface tension and elasticity, $\glv^{2/3} Y^{1/3}$, and so we expect the tension in the sheet to be significantly modified by the drop's presence.

Recalling the definition (\ref{eq:tauinf}) of the dimensionless parameter $\tauinf$, one may note that the above classification into perturbative and non-perturbative responses (now based on the nature of the solutions of the FvK equations), mirrors precisely the distinction made in \S2.1 on the basis of energy considerations. 

\section{Results\label{sec:results}}

We now report the results of our solution of the FvK equations, as formulated in \S\ref{sec:details} (with further details given  in Appendix A). In \S\ref{sec:contact} we focus on the contact geometry and stress in the vicinity of the contact line. To simplify this problem as much as possible, we first consider an infinitely large sheet, $\Rout/\Rin=\infty$, so that the details of the boundary conditions (clamped versus load-controlled) are not significant apart from the value of the far-field tension, $\Tinf$, that is imposed. We then explain how our solution can be used to properly extract the value of $\Tinf$ from the measured  angle $\phi$, and identify the parameter regime in which the measurement of the tension at the contact line, reported 
by Nadermann \emph{et al.}~\cite{Nadermann13} and Schulman \& Dalnoki-Veress~\cite{Schulman15}, gives a good estimate of $\Tinf$. 
In \S\ref{sec:finitesize} we discuss the conditions under which our neglect of the finite size of the system is valid, and discuss the differences between clamping and load-controlled conditions. Finally, in \S\ref{sec:phasediag} we discuss  the ``phase diagram" (fig.~2) that characterizes the parameter regimes in which the capillary-induced stress is weak (perturbative) or strong (non-perturbative) when compared to the pre-existing stress in the sheet. We illustrate the differences between these regimes by discussing the nature of the stress profiles induced by the presence of the drop.

\subsection{Stress and angle in the vicinity of the  contact line\label{sec:contact}}

\subsubsection{Solution of FvK equations}

\begin{figure}
\centering
\includegraphics[width=0.8\columnwidth]{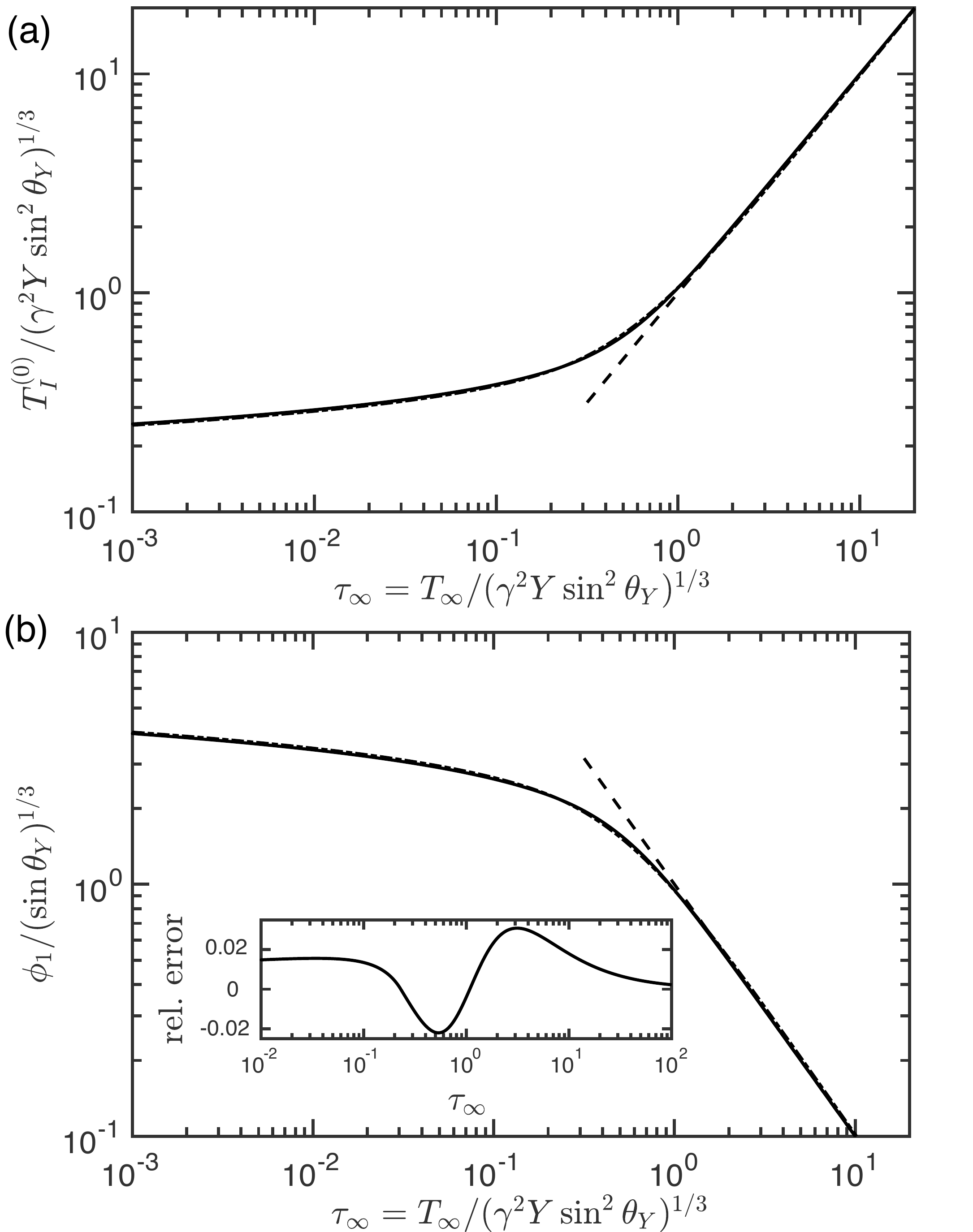}
\caption{ The numerically-determined results of the leading order problem. (a) The rescaled leading-order tension at the contact line, $\Ti^{(0)}/(\glv^2Y\sin^2\theta_Y)^{1/3}$, as a function of the imposed load, $\btauinf=\tauinf/\sin^{2/3}\theta_Y$. Plotted in this way, the numerical results (solid curve) are universal in the singular limit $\tglv\to0$. For $\btauinf\to\infty$, $\Ti\approx \Tinf$ (black dashed line) and the tension at the contact line is only slightly perturbed by the presence of the drop. However, as $\btauinf$ becomes order unity, we find a significant perturbation caused by the drop.  (b) The rescaled angle of inclination of the drop, $\phi_1\approx\phi/(\glv/Y)^{1/3}$, as a function of the applied tension, $\btauinf$. Numerical results are shown (solid curve), together with the approximate expression \eqref{eqn:PhiAnalytical} (dash-dotted curve) and the large $\btauinf$ limit $\phi_1\approx \sin\theta_Y/\Tinf$ (dashed line). Inset: The relative error in \eqref{eqn:PhiAnalytical} as $\btauinf$ varies; this error remains less than $3\%$ throughout.  Note that the two quantities $\Ti^{(0)}$ and $\phi_1$ are `slaved' by the leading order vertical force balance at the contact line, \eqref{eq:phi1explicit},  so that the product of the quantities on the $y$-axes of (a) and (b) is unity.}
\label{fig:Universal}
\end{figure}


\begin{figure}
\centering
\includegraphics[width=0.8\columnwidth]{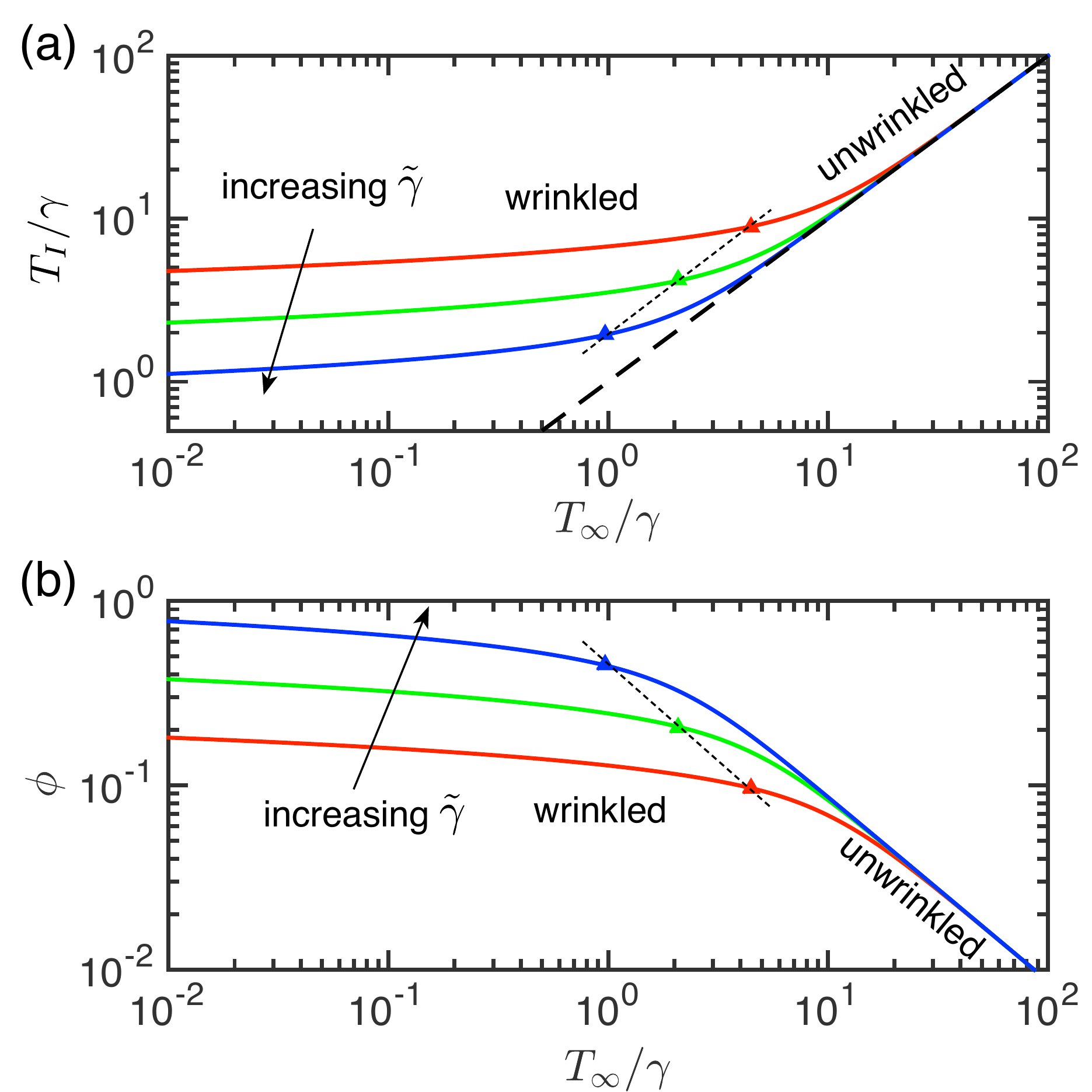}
\caption{ The effect of the transition from perturbative to non-perturbative behaviour on measurable quantities for different values of the capillary  extensibility $\tglv$: $\tglv=10^{-4}$ (red), $10^{-3}$ (green) and $10^{-2}$ (blue). (a) The tension on the 
wet side of the contact line $\Ti$ and (b) the inclination angle of the sheet on the wet side of the contact line, $\phi$. In both plots, results are shown for an infinite sheet, $\Rout=\infty$, (so that clamping and applied load boundary conditions are identical), and $\theta_Y=60^\circ$. Plotting in this  way highlights that the degree of pre-tension, $\Tinf/\glv$, required for the presence of the droplet to be perturbative depends on the value of $\tglv$. The boundary between wrinkled and non-wrinkled states is indicated by triangles and a dashed line.}
\label{fig:FiniteTgamma}
\end{figure}

We solved the FvK equations numerically (following the procedure described in Appendix A) for a partially wet sheet, which is large (i.e.~$\Rout/\Rin\gg1$), and nearly inextensible when subject to capillary-induced loads (i.e.~$\tglv=\glv/Y \ll 1$). In this regime the physics is governed by a {\emph{single}} dimensionless parameter:  
\begin{equation}
\btauinf=\frac{\tauinf}{ \sin^{2/3}\theta_Y}=T_\infty/(\glv^2Y\sin^2\theta_Y)^{1/3} \ . 
\label{eq:define-tauinf-2}
\end{equation}

Since a direct measurement of stresses within a solid sheet is difficult, 
a primary purpose of experiments is to use measurements of the inclination angle $\phi$ 
to infer the stress, $\Ti$, at the contact line (see fig.~\ref{fig:setup}), and hence the tension in the dry sheet $\Tdry$. Realizing that our solution of the FvK equations provides the shape of the deformed sheet and the stress within it as functions of the single parameter, $\btauinf$, we can extract from this solution a direct relationship between $\phi$ and $\Tinf\approx\Tdry$, with which an experimentalist can directly infer the latter by measuring the former. 
Since our focus is on nearly inextensible sheets, namely, $\tglv \ll 1$, we expand these quantities in powers of $\tglv^{1/3}$, so that
\beq
\frac{\Ti}{\glv^{2/3}Y^{1/3}}=\bTi=\bTi^{(0)}+O(\tglv^{1/3}),\quad \phi=\phi_1\tglv^{1/3}+O(\tglv^{2/3}),
\eeq with the exponent $1/3$ rationalized in Appendix A.  

Our numerical predictions for the leading order functions $\bTi^{(0)}$ and $\phi_1$ are shown in fig.~\ref{fig:Universal}a and \ref{fig:Universal}b, as the single dimensionless parameter $\btauinf$ varies. (Note that at leading order the inclination angle $\phi$ is ``slaved" to the tension at the contact line, $\Ti$, through the relationship, $\phi \approx \sin\theta_Y \times \gamma/\Ti$, as  one would anticipate from a simple vertical force balance. At leading order (in $\tglv$), this slaving is exact, i.e.~$\phi_1=\sin\theta_Y/\bTi^{(0)}$, and so the curves in figs~\ref{fig:Universal}a and \ref{fig:Universal}b are the reciprocal of one another.) The analogous panels in fig.~\ref{fig:FiniteTgamma} show $\Ti$ and $\phi$ as functions of $\Tinf/\gamma$ for a few values of the capillary extensibility parameter, $\tglv$. 
\\

Figure~\ref{fig:Universal}a exhibits a clear division into two regimes. For $\btauinf > O(1)$, $\Ti \approx \Tinf$ (dashed line); in contrast, for $\btauinf < O(1)$, $\Ti \gg \Tinf$. These distinct types of mechanical response parallel our qualitative discussion of \S\ref{sec:qualitative}: For sufficiently large values of  $\btauinf$, the capillary-induced stress is weak compared to the pre-tension and hence $\Ti \approx \Tinf$. However,  for small values of $\btauinf$ the capillary-induced stress is strong compared to the pre-tension and therefore is responsible for the value of $\Ti$. One signature of the significant change to the stress seen for `weak' pre-tension is the onset of radial wrinkles when the hoop stress becomes compressive in an annular zone around the contact line, \emph{i.e.}~$\sqq(r) <0$ for $r \approx \Rin$;  we find that wrinkling occurs for $\btauinf\lesssim0.2297$.

Importantly, the borderline between ``strong" and ``weak" pre-tension {\emph{is not}}, $\Tinf \sim \gamma$, as one might naively guess, but rather, $\btauinf \sim O(1) \Rightarrow \Tinf \sim T^\ast = \gamma^{2/3} Y^{1/3} \gg \gamma$. While this prediction was already anticipated through the qualitative arguments of \S\ref{sec:qualitative}, figs \ref{fig:Universal}  and \ref{fig:FiniteTgamma} show that in the non-perturbative regime, $\Tinf < T^\ast$, the value of $\Ti$ is not constant; instead, as we discuss below, our solution to the FvK equations exhibits a weak, logarithmic dependence of $\Ti$ on $\Tinf$ in the non-perturbative regime.    
\\

To facilitate the use of our numerical results in data analysis, it would be helpful to provide an explicit approximant for the curve in fig.~\ref{fig:Universal}b, so that an experimenter can extract the pre-tension in the sheet 
($\Tinf$) from measurements of either the inclination angle of the sheet at the contact line, $\phi$, or the measured tension in the vicinity of the contact line, $\Ti$, without needing to  solve the FvK equations numerically. It is possible to determine an asymptotic result for $\phi$ in the limit $\btauinf\ll1$, as shown in Appendix \ref{sec:theory} (see, in particular, eqn \eqref{eqn:AppATin}). Similarly, for $\btauinf\gg1$, we expect to recover \eqref{eq:approachYoung-2}, or, equivalently, $\phi\sim (\tglv\sin\theta_Y)^{1/3}\btauinf^{-1}$. For simplicity, we suggest that the asymptotic result for $\btauinf\ll1$ be used whenever $\btauinf\lesssim0.2297$ (\emph{i.e.}~whenever wrinkling occurs). For $\btauinf\gtrsim0.2297$ (with no wrinkling), we suggest using a functional form $\phi=(\tglv\sin\theta_Y)^{1/3}\bigl[(\btauinf+A_3)/(\btauinf^2+A_1\btauinf+A_2)\bigr]$ which reproduces the expected behaviour for $\btauinf\gg1$; here, the constants $A_1\approx0.839$, $A_2\approx0.351$ and $A_3\approx1.069$ are  chosen so that this expression joins smoothly to the result for $\btauinf\lesssim0.2297$ at this critical point. 
This yields a suitable approximation for the membrane inclination (measured in radians): 
\begin{equation}
\frac{\phi}{(\glv/Y)^{1/3}\sin^{1/3}\theta_Y}
\approx   \begin{cases} (C-10\log\btauinf)^{1/3}, \quad \btauinf\lesssim 0.2297\\
\frac{\btauinf+A_3}{\btauinf^2+A_1\btauinf+A_2},\quad \btauinf\geq 0.2297.
\end{cases}
\label{eqn:PhiAnalytical}
\end{equation}  while the internal tension at the contact line is simply given by $\Ti\approx\glv\sin\theta_Y/\phi$. In eqn \eqref{eqn:PhiAnalytical}, the constant $C\approx-4.394$ emerges from a detailed asymptotic analysis of the wrinkled problem (see Appendix A) and the logarithm is natural. 
The expression in \eqref{eqn:PhiAnalytical} is accurate to within $3\%$ for all $\btauinf$ (see inset of fig.~\ref{fig:Universal}(b)).

\subsubsection{Extracting $\Tinf$ from measured angles\label{sec:extract}} 

In the experimental studies of Nadermann \emph{et al.}~\cite{Nadermann13} and Schulman \& Dalnoki-Veress~\cite{Schulman15}, the authors measured the angles at the contact line of a droplet sitting on a suspended sheet. From these measurements, they extracted the value of the stresses $\Ti$ and $\To$ using force balance at the contact line. In \S\ref{sec:pretension} and \S\ref{sec:angles} we address some subtleties in both versions of this proposal that appear not to have been appreciated previously. Here we focus instead on a separate assumption, which we believe was made implicitly by both groups of authors: that the extracted tension ($\Ti$ or $\To$) gives a reliable estimate for the tension within the sheet prior to wetting, $\Tdry\!\approx\!\Tinf$ \footnote{ 
We emphasize that in this paper $\Ti,\To$, and $\Tinf$, refer to values of the tension in the sheet in the vicinity of the contact line ($\Ti,\To$) and away from it ($\Tinf\approx\Tdry$) for a \emph{given} sheet thickness $t$. Our results are thus akin to an individual experiment in Fig.~3 of Ref.~\cite{Nadermann13} and Fig.~2 of Ref.~\cite{Schulman15}), and \emph{not} to any extrapolations of $\Ti(t),\To(t), \Tinf(t)$ as $t\to 0$, which are shown in Tables 1-3 of Ref.~\cite{Nadermann13}. We do not discuss here the possible meaning of such an extrapolation, which was proposed by Nadermann \emph{et al.}~\cite{Nadermann13} as a means to determine solid surface energies.}.

\begin{figure}
\centering
\includegraphics[width=0.8\columnwidth]{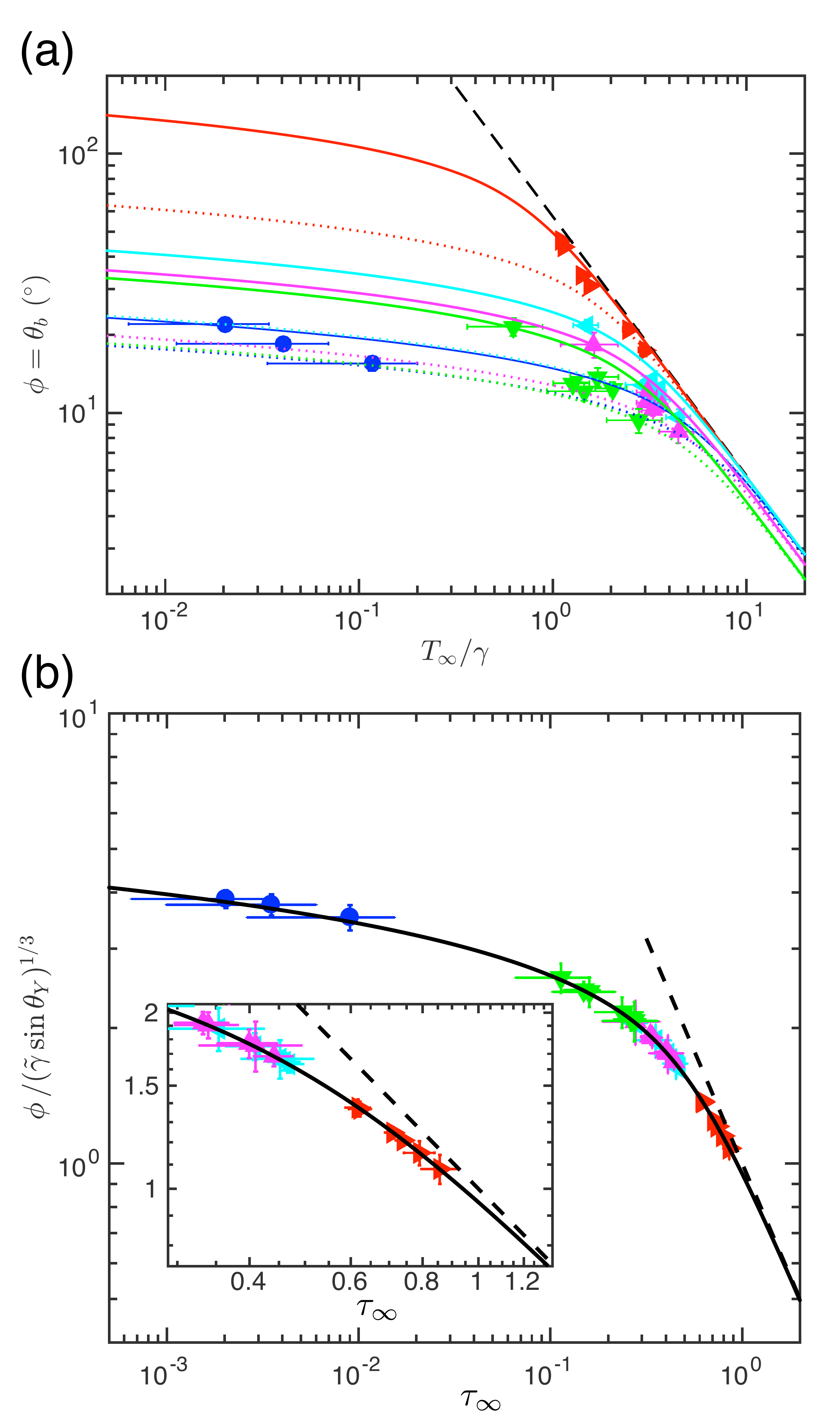}
\caption{ Analysis of previously published experimental measurements of the membrane angle $\phi$ for different droplets on suspended elastic membranes of different thickness and material properties \cite{Nadermann13,Schulman15}. (a) Measurements of $\phi$ may be used to infer the appropriate value of $\Tinf/\gamma$ according to the leading order theory presented here. (b) Rescaling the membrane angle $\phi$ in the expected way collapses data onto the theoretical prediction (solid curve). The deviation of the data from the perturbative prediction, $\phi=\tglv^{1/3}\sin^{1/3}\theta_Y/\btauinf$, (dashed line) shows the extent to which experiments fall into the perturbative regime. The inset highlights the region in which  the transition from perturbative to non-perturbative occurs.  In all plots circles represent data   for PnBMA films \cite{Schulman15} (the large error bars in the non-perturbative regime stem from the weak (logarithmic) dependence of $\Ti$ and $\phi$ on the pre-tension). Right pointing triangles show data for SIS \cite{Schulman15} . Other triangles show  results  for PDMS sheets \cite{Nadermann13} with different liquid droplets as follows: De-ionized water (left pointing triangles), Ethylene-Glycol (upward pointing triangles) and DMSO (downward pointing triangles). In the calculation of the associated pre-tension, $\Tinf$, we use values of the Young moduli as follows: $E=0.8\mathrm{~MPa}$ (SIS, as communicated by Kari Dalnoki-Veress), $E=1\mathrm{~GPa}$ (PnBMA, as communicated by Kari Dalnoki-Veress) and  $E=2.85\mathrm{~MPa}$ (PDMS, based on the value used for numerical simulations by Nadermann \emph{et al.}~\cite{Nadermann13}).}   
\label{fig:Comparison-1}
\end{figure}

At this stage of our discussion, it should be clear that the validity of the approximation, $\Tinf \approx \Ti$, is limited to the parameter regime, $\Tinf \gg T^\ast \sim \gamma^{2/3}Y^{1/3}$, for which the capillary-induced stress is sufficiently weak. The results of our FvK model (to leading order in $\tglv$)  allows us to extract the actual far-field tension $\Tinf$, that would lead to a given inclination angle, $\phi$ --- if the values of the sheet's thickness and Young's modulus ($t,E$, respectively), the liquid--vapour surface tension ($\gamma$), and the Young angle ($\theta_Y$) are all known, then the inclination angle $\phi$ is enough to determine $\Tinf$. The measured values of this angle for several systems were kindly provided to us by  Nadermann \emph{et al.}~\cite{Nadermann13} and Schulman \& Dalnoki-Veress~\cite{Schulman15}; fig.~\ref{fig:Comparison-1} shows our determination of the corresponding values of $\Tinf$ in each case, based solely on the reported measurement of the angle $\phi$ and the material properties of the sheets involved\footnote{Note that fig.\ref{fig:Comparison-1} shows how to employ FvK theory to extract a far-field tension, $\btauinf$, from measured angles in the absence of \emph{a priori} knowledge of this tension; it should \emph{not} be interpreted as an experimental validation of our theory.}.  Table \ref{table:1} reports each set of data and compares the value of $\Tinf/\glv$ determined via our method with the corresponding value of $\Ti/\glv$, 
which were previously reported \cite{Schulman15,Nadermann13}.  Two observations are immediately apparent from fig.~\ref{fig:Comparison-1}: 

$\bullet$ For several systems (specifically, drops of glycerol on SIS \cite{Schulman15}), the approximation $\Ti \approx \Tinf$ gives a good estimate of the true tension within the sheet. However, for glycerol drops on PnBMA \cite{Schulman15} (the blue circles in fig.~\ref{fig:Comparison-1}), the value of $\Ti$ exceeds our estimate of the actual $\Tinf$ by more than an order of magnitude. Furthermore, for some of the thinner PDMS sheets (e.g.~the sheets with $t=5-15\mathrm{~\mu m}$ and drops of De-ionized water or DMSO, see table \ref{table:1}) there can be a smaller, but still substantial difference, $O(50\%)$ or more, between the perturbative approach \cite{Nadermann13} and the FvK-based solution. (Note that two values of $\Tinf$ are given for PDMS in table \ref{table:1},  corresponding to the range of values of Young modulus, $0.57\mathrm{~MPa}\leq E \leq3.7\mathrm{~MPa}$ \cite{Wang14}; for a given measured angle $\phi$, the larger value  of $E$ leads to larger discrepancies between our FvK-based calculations and the perturbative approach.)

The question then arises of whether one can know which of the two regimes (perturbative or non-perturbative) an experiment lies in. Without knowing the  tension in the dry sheet (denoted $\Tdry$ and defined in \eqref{eq:defineTdry} dependent on the boundary conditions)  such knowledge is not possible.
However, once the angle $\phi$ has been measured (and assuming the values of $\tglv$ and $\theta_Y$ are known) it is possible to test whether the droplet lies in the perturbative or non-perturbative regime. Our numerical solution shows that if the measured value of $\phi$ (in radians) satisfies
\beq
\phi\lesssim 1.18 \left(\frac{\glv\sin\theta_Y}{Y}\right)^{1/3},
\label{eqn:ruleofthumb}
\eeq then the value of $\Ti$ (inferred from vertical force balance) is within $10\%$ of the value of the far-field tension $\Tinf$. We 
propose \eqref{eqn:ruleofthumb} as a rule-of-thumb to determine whether experiments lie in the perturbative or non-perturbative regimes; 
we used this criterion in potting 
the perturbative and non-perturbative regions in fig.~\ref{fig:RegimeDiagSlivers}.

$\bullet$ In the non-perturbative regime, the weak (logarithmic) dependence of $\Ti$ (and consequently the inclination angle $\phi$) on the pre-tension, gives rise to large ``signal-to-noise" ratio. Hence, even an extremely precise measurement 
({\emph{e.g.}~2\% accuracy), gives rise to a large uncertainty in inferring 
$\Tinf$ 
from the value of $\phi$.   
One should bear this sensitivity in mind when using ``drop-on-sheet" as a set-up for determining pre-tension in suspended sheets.
\\

\begin{table*}[h]
\small
  \caption{A summary of previous experiments on sheets of SIS and PnBMA (data taken from Schulman \& Dalnoki-Veress~\cite{Schulman15}) and PDMS (data taken from Nadermann \emph{et al.}~\cite{Nadermann13}). The raw values of the measured angle $\phi$ in our notation are given, together with the calculated tension at the contact line, $\Ti$. We also give the calculated value of the far-field tension $\Tinf$ based on the analysis presented in this paper. Values for PnBMA (SIS) use the estimate $E=1\mathrm{~GPa}$ ($E=0.8\mathrm{~MPa}$), as communicated to us by Kari Dalnoki-Veress. The value of Young's modulus for PDMS is known \cite{Wang14} to vary in the range $0.57\mathrm{~MPa}\leq E\leq3.7\mathrm{~MPa}$. Since Nadermann \emph{et al.}~\cite{Nadermann13} do not include values of $E$ for the samples used in their experiments, we calculated the pre-tension $\Tinf$ that correspond to the measured angles for each sheet thickness using the extreme values given by Wang \emph{et al.}~\cite{Wang14} as lower and upper bounds of $E$,  these correspond to the upper and lower bounds of $\Tinf$, respectively.}
  \label{table:1}
  \begin{tabular*}{\textwidth}{@{\extracolsep{\fill}}cccccccc}
Sheet &  $t\mathrm{~(\mu\,m)}$ & Liquid & $\lbc \mathrm{~(\mu m)}$ & $\phi~(^\circ)$ & $\Ti/\glv$    & \multicolumn{2}{c}{$\Tinf/\glv$} \\
\hline
PnBMA & 0.065 & Glycerol  & 0.64 & 22  & 2.50& \multicolumn{2}{c}{0.020}\\
PnBMA & 0.1 &  Glycerol & 1.21 & 18.5 &2.99& \multicolumn{2}{c}{0.041}\\
PnBMA & 0.14 & Glycerol & 2.01 & 15.5 & 3.61 & \multicolumn{2}{c}{0.12}\\   
\hline 
SIS & 0.27 & Glycerol  & 0.17 & 46  &1.20& \multicolumn{2}{c}{1.10} \\
SIS &0.35 &  Glycerol & 0.24 & 43.5 & 1.26& \multicolumn{2}{c}{1.15}\\
SIS & 0.97 & Glycerol & 1.13 & 34 & 1.62 & \multicolumn{2}{c}{1.41}\\    
SIS & 1.3 & Glycerol & 1.75 & 31 &1.77 & \multicolumn{2}{c}{1.55}\\    
SIS & 2.45 & Glycerol & 4.52 & 21 & 2.61 & \multicolumn{2}{c}{2.45}\\
SIS & 3.5 & Glycerol & 7.72 & 17.5 & 3.15 & \multicolumn{2}{c}{2.99}\\   
\hline 
& &   &  & & & $\Tinf/\gamma$  & $\Tinf/\gamma$ \\
& &   &  & & & ($E=3.7\mathrm{~MPa}$) & ($E=0.57\mathrm{~MPa}$)\\
PDMS & 5 & De-ionized Water & $7.5-19.0$ & 21.8 & 2.6 & $1.2$ & $2.4$\\    
PDMS & 9 & De-ionized Water & $25.2-64.2$& 13.3 & 4.3 & $3.0$ & $4.1$\\    
PDMS & 11 & De-ionized Water & $34.0-86.7$ & 12.4 & 4.7 & $3.2$ & $4.4$\\    
PDMS & 15& De-ionized Water & $54.2-138.1$ & 12.8 & 4.5 & $2.6$ & $4.1$\\    
PDMS & 18 & De-ionized Water & $71.2-181.5$& 11.3 & 5.1& $3.3$ & $4.8$ \\    
PDMS & 25 & De-ionized Water & $116.6-297.0$ & 9.6 & 6.0 & $4.1$ & $5.6$ \\    
\hline
PDMS &5 & Ethylene Glycol &$9.3-23.8$ & 18.3 & 2.6 & $1.4$ & $2.6$\\    
PDMS & 9& Ethylene Glycol & $31.5-80.3$ & 12.0 & 4.0 & $2.8$ & $4.1$\\    
PDMS & 11& Ethylene Glycol & $42.6-108.5$ & 10.7 & 4.5 & $3.3$ & $4.6$\\    
PDMS & 15& Ethylene Glycol & $67.8-172.7$& 11.0 & 4.5 & $2.6$ & $4.4$\\    
PDMS & 18& Ethylene Glycol & $89.1-227.1$& 10.2 & 4.9 & $2.8$ & $4.7$\\ 
PDMS & 25& Ethylene Glycol & $145.9-371.7$ & 8.5 & 6.0 & $4.0$ & $5.8$\\ 
\hline   
PDMS & 5& DMSO & $9.7-24.6$& 21.5 & 2.3 & $0.4$ & $1.7$\\    
PDMS & 9& DMSO & $32.6-83.0$ & 13.8 & 3.5 & $1.4$ & $3.0$\\    
PDMS & 11& DMSO & $44.0-112.2$& 12.4 & 3.9 & $1.6$ & $ 3.3$\\    
PDMS & 15& DMSO & $70.1-178.6$ & 13.0 & 3.8 & $0.9$ & $3.0$\\    
PDMS & 18& DMSO & $92.1-234.8$& 12.1 & 4.1 & $1.0$ & $3.2$\\    
PDMS & 25& DMSO & $150.8-384.3$ & 9.4 & 5.4 & $2.3$ & $4.4$\\ 
    \hline
  \end{tabular*}
\end{table*}

\subsection{
Finite size and boundary conditions  \label{sec:finitesize}}

In the analysis presented above, it has been assumed that the sheet can be considered infinitely large in comparison to the drop's radius, such that a uniform, isotropic stress, $\srr = \sqq = \Tinf$ is attained sufficiently far from the contact line. In practical situations, one has to relate the effective far-field tension $\Tinf$ that appears in eqn \eqref{eqn:PhiAnalytical} and figs \ref{fig:Universal},\ref{fig:FiniteTgamma} to the actual boundary condition imposed at 
$r=\Rout$. The two most common boundary conditions in experiments are a fixed tensile load $\Tedge$ or an imposed horizontal displacement (corresponding to a pre-tension, $\Tpre$, with subsequently clamped boundary condition). These two tensions are both specific cases of the tension in the dry sheet,  $\Tdry$, which we defined in \eqref{eq:defineTdry}. 
In the first part of this subsection, we consider briefly the different mechanisms through which $\Tinf$ is determined under these two types of BCs. We will see that in both cases, one can safely ignore any effect of finite size (so that $\Tinf \approx \Tdry$) for practical purposes, provided that the sheet is sufficiently large, $\Rout/\Rin \gtrsim 10$. In the second part, we briefly address the differences we would expect to observe in the elasto-capillary response of multiple drops on a suspended sheet, in comparison to the single drop model assumed throughout our study.

\subsubsection{Boundary conditions}
\emph{Prescribed load:} This has been achieved experimentally by using sheets floating on a liquid bath  \cite{Huang07,Schroll13,Toga13,Boulogne16}, and variants thereon \cite{King12,Pineirua13,Paulsen17}. In this case, $\srr=\Tedge$ is prescribed at $r=\Rout$; using the axisymmetric Lam\'{e} solution \eqref{eq:Lameaxi} gives 
\begin{equation}
\Tedge=\Tinf+(\To-\Tinf)\frac{\Rin^2}{\Rout^2}.
\label{eq:Tedge-1}
\end{equation} Note that $\To$ is a function of $\Tinf$ (at this order in our asymptotic expansion $\To=\Ti$, with $\bTi(\btauinf)$ plotted in fig.~\ref{fig:Universal}a). We see therefore that \eqref{eq:Tedge-1} gives us an implicit equation for $\Tinf(\Tedge)$.  

\emph{Clamping with pre-tension:} If the sheet is initially subject to some (isotropic) pre-tension, $\Tpre$, and then clamped in this state then the horizontal displacement at the edge is fixed: $u_r(\Rout)=(1-\nu)\Tpre\Rout/Y$. Any subsequent stress field must also satisfy this condition on the radial displacement at $r=\Rout$; the (axisymmetric)  Lam\'e  solution \eqref{eq:Lameaxi} then gives that:
\begin{equation}
\Tpre=\Tinf-\frac{1+\nu}{1-\nu}(\To-\Tinf)\frac{\Rin^2}{\Rout^2}.
\label{eq:Tedge-2}
\end{equation}
\\

\begin{figure}
\centering
\includegraphics[width=0.8\columnwidth]{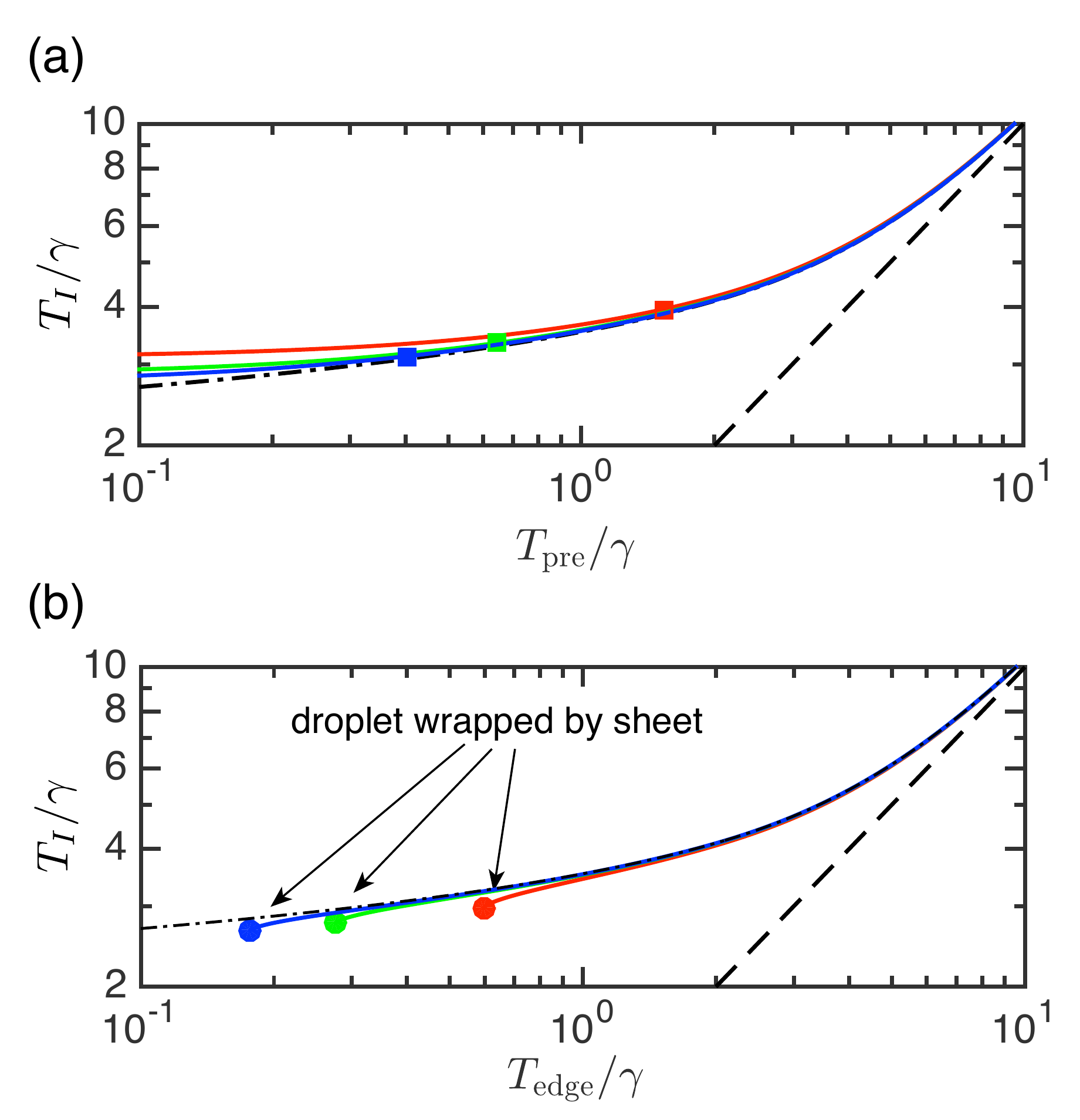}
\caption{The effect of finite size (hence different boundary conditions) on the leading order tension at the contact line, $\Ti^{(0)}=\To^{(0)}$. Solid curves show numerical results computed with various sheet sizes: $\Rout/\Rin=5$ (red), $10$ (green) and $15$ (blue).
The corresponding results for an infinite sheet are shown for comparison (dash-dotted curve), together with the perturbative prediction, $\Ti=\Tinf$ (dashed line). (a) Results for a  sheet clamped while under a pre-tension $\Tpre$. Squares indicate where the value of $\Tpre$ for a given $\Ti$ differs by $10\%$ from the infinite sheet result.
(b) The case of a controlled load $\Tedge$ applied at $r=\Rout$. In this case, wrinkles ultimately reach the edge of the sheet, and the droplet is wrapped by the sheet if the applied tension falls below a critical value, indicated by circles. In both plots $\tglv=10^{-3}$
and $\theta_Y = 60^\circ$.}
\label{fig:FiniteSize}
\end{figure}

Equations~\eqref{eq:Tedge-1} and \eqref{eq:Tedge-2} suggest that, as long as $\Rout/\Rin \gg 1$, 
the far-field stress $\Tinf$ is very close to the tension within the dry sheet \emph{i.e.}~$\Tinf\approx\Tdry$,  up to corrections of $O(\Rin/\Rout)^2$. From our numerical solutions of the FvK equations (see figure \ref{fig:FiniteSize}), we find that these approximations are highly accurate as long as $\Rout/\Rin \gtrsim 10$, and $\Tdry$ is not too small. For practical purposes, therefore, the two types of boundary conditions (clamped and load-controlled) are likely to be almost indistinguishable, implying that $\Tdry \approx \Tinf$.  

This result comes with a proviso, however: the  axisymmetric results \eqref{eq:Tedge-1} and \eqref{eq:Tedge-2} also demonstrate that when $\To$ becomes sufficiently large in comparison to the dry tension  $\Tdry$ (which may occur only when $\tauinf$ is extremely small and the system is deeply in the non-perturbative regime) a better estimate is required. Indeed, fig.~\ref{fig:FiniteSize}a shows that even with $\Rout/\Rin=15$, the effect of finite size may be observed once $\Tpre/\glv\lesssim0.1$. In this `ultra-weak tension' regime, one needs to be careful to account for wrinkling (which occurs with $\btauinf\lesssim0.2297$); careful consideration shows that the effect of finite size become relevant when $\Rout/\Rin\sim \To/\Tinf$, rather than the critical scale $\Rout/\Rin\sim (\To/\Tinf)^{1/2}$ that might be expected from \eqref{eq:Tedge-1},\eqref{eq:Tedge-2}. Since $\To\sim\glv^{2/3}Y^{1/3}$ in the non-perturbative regime (up to logarithmic corrections) we find that the finite size of the system, and hence the details of the boundary conditions become important when
\begin{equation}
\Tdry\lesssim \gamma^{2/3}Y^{1/3} \frac{\Rin}{\Rout} .
\label{eq:min-T}
\end{equation}

For `ultra-weak' tensions, \emph{i.e.}~those satisfying \eqref{eq:min-T}, the two types of BCs give rise to qualitatively different responses. If the edge is clamped, the sheet remains nearly planar, but the far-field tension, 
$\Tinf$,  is far larger than the pre-tension, $\Tpre$ (see fig.~\ref{fig:FiniteSize}a and \ref{fig:Profiles}a). If the sheet is subject to a controlled load, with $\Tedge$ satisfying \eqref{eq:min-T}, the drop ``pulls" the whole sheet, and a transition from partial wetting to a complete wrapping is expected; such a mechanism underlies the wrapping of liquid drops by sheets and the geometry-induced transition from wrinkles to folds in floating sheets, both of which have been reported  recently \cite{Paulsen15, Paulsen17}.  

\subsubsection{Placing multiple drops on a sheet}
An interesting question, triggered by various experiments, pertains to the effect of placing multiple drops on a suspended  \cite{Schulman15} or floating  \cite{HuangThesis} sheet. Clearly, if the system is in the perturbative regime, where the capillary--induced stress is weak in comparison to the stress at the far edge, there is little interaction between the drops, even if their mutual distance is comparable to or even smaller than a drop's radius. However, in the non-perturbative regime, where each drop induces a non-uniform, anisotropic stress around itself (see dashed green curves in Fig.~\ref{fig:Profiles}), one may expect a mutual interaction between drops; this may affect a net, density-dependent contribution to the far-field stress. Since the capillary-induced stress decays with distance from the contact line (unless the system is in the ultra-weak tension regime, defined by Eq.~\ref{eq:min-T}), we expect the mechanics to still be described by our single-drop model if the density of drops is sufficiently small. A simple estimate for the minimal density of drops, $\rho_{\mathrm{min}}$, above which drop-drop interactions prevail \footnote{At the qualitative level, such a transition between weak and strong interaction regimes was explored in chapter 4 of Huang \cite{HuangThesis}. There, multiple drops were placed on a floating sheet and substantial drop-drop interactions (evidenced through wrinkle patterns of various shapes), were observed when the drop separation was comparable to their radii.}, may be obtained by noting that  
interactions must certainly occur 
when the wrinkle patterns of isolated drops start to overlap. Since the wrinkle length $L\sim \To\Rin/\Tinf$, \cite{Schroll13} we expect that: 
\begin{equation}
\rho_{\mathrm{min}}\lesssim (\Rin/L)^2\sim (\Tinf/\To)^2\sim \btauinf^2
\label{eq:rho_min}
\end{equation}

\subsection{A ``phase diagram" for a partially wet solid sheet\label{sec:phasediag}}

\begin{figure}
\centering
\includegraphics[width=0.8\columnwidth]{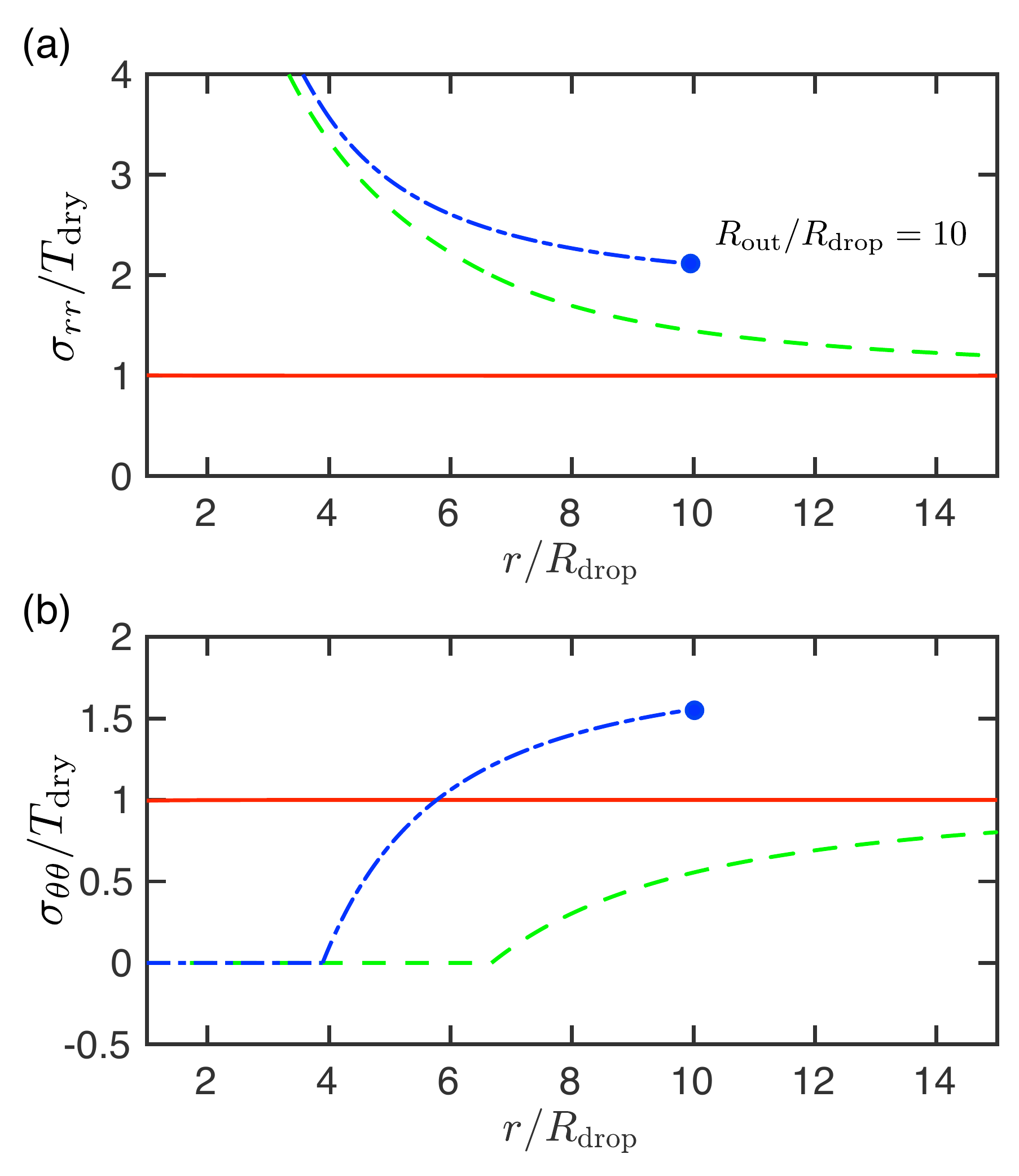}
\caption{Stress profiles within the sheet as the 
pre-tension and system size vary. Here results are shown for $\Tdry/\glv=10$ (red, solid curves) and $\Tdry/\glv=0.1$ (green, dashed curves and blue, dash-dotted curves), all with $\theta_Y=60^\circ$ and $\tglv=10^{-2}$. Results for $\Tdry/\glv=0.1$ are shown for an infinite system (green, dashed curves) and a clamped system with $\Rout/\Rin=10$ (blue, dash-dotted curves). (a)  The radial stress, $\srr$, and (b) hoop stress, $\sqq$, are shown. We note that the stress profile within the sheet is highly anisotropic, $\srr\neq\sqq$, unless $\Tdry/\glv$ is large. Furthermore, the hoop stress becomes relaxed, $\sqq\approx 0$, for some region in response to the appearance of radial wrinkles. Finally, note that the combination of a finite clamped boundary and ultra-low tension means that the stress within the sheet is significantly perturbed from the value of the pre-tension everywhere within the sheet,  not merely in the vicinity of the droplet (compare the finite case with $\Tdry/\glv=0.1$ to that  with the same value of $\Tdry$ but an infinite system size).
}
\label{fig:Profiles}
\end{figure}

Figure~\ref{fig:RegimeDiagSlivers} provides a succinct summary of our results in terms of the different behaviours of the system that are observed in different regimes of the capillary and dry extensibility parameters, $\tglv$ and $\tTdry$ (defined in \eqref{eq:define-dry-exten}). Figure~\ref{fig:RegimeDiagSlivers} is thus a ``phase diagram" for partial wetting of a large ($\Rout \gg \Rin$), highly bendable  ($\epscap \ll 1$) sheet. The diagram distinguishes between three asymptotic regimes: 
``perturbative", ``non-perturbative", and ``ultra-weak tension" (where the tension is weak enough that finite size effects, \emph{e.g.}~the precise boundary condition applied at $r=\Rout$, plays an important role). The qualitative differences between these ``wettability phases" stem from the distinct mechanism by which the capillary-induced stress affects the isotropic, uniform stress, $\srr = \sqq = \Tdry$  in the sheet prior to wetting. 
\\

\underline{\emph{Perturbative regime}, $\tauinf \gg 1$:} (Equivalent to: $\tTdry \gg \tglv^{2/3}$)

\noindent In this parameter regime, the stress in the partially-wet sheet is only very slightly  perturbed from the isotropic, uniform stress of the dry sheet. This is demonstrated in fig.~\ref{fig:Profiles}, where the solid red curves show the stress components, $\srr(r)$ (fig.~\ref{fig:Profiles}a) and $\sqq(r)$ (fig.~\ref{fig:Profiles}b) for a sheet with $\tTdry=0.1$ and $\tglv = 0.01$.

The implications of this perturbative effect on the vicinity of the contact line are rather obvious --- indeed, the deviations of the solid red curves from the uniform tension state are barely visible on the scale of fig.~\ref{fig:Profiles}. In the vicinity of the contact line, the stress $\To = \srr(\Rin^+) \approx \Tinf$,  and the stress discontinuity (which, to leading order in $\tglv$, is given by the YLD contact mechanics (\ref{eq:YLDmechanics-define}) and is unaffected by elasticity, see \S\ref{sec:contactline}) implies $\Ti = \srr(\Rin^-) \approx \Tinf +\gamma \cos\theta_Y \approx \Tinf$ (where the last equality follows since $\Tinf \gg T^\ast\gg \gamma$).
\\

\underline{\emph{Non-perturbative regime}, $ \Rin/\Rout<\tauinf\ll1$} (Equivalent to $ \tglv^{2/3}\Rin/\Rout < \tTdry \ll \tglv^{2/3}$)

\noindent In this parameter regime, the stress in the partially-wet sheet exhibits an anisotropic and non-uniform profile around the drop. The stress deviates strongly from the uniform, homogeneous stress state prior to wetting, as is demonstrated in fig.~\ref{fig:Profiles} for a sheet with $\tTdry=0.001$ and $\tglv = 0.01$ (dashed green curves). In this case, both stress components return to their dry states far from the droplet, i.e.~$\srr,\sqq\to\Tdry$ as $r/\Rin\to\infty$. However, each stress component demonstrates a significant change in the vicinity of the drop: firstly, the hoop stress vanishes, $\sqq=0$, in some region that spans the contact line. (This plateau heralds the formation of radial wrinkles, which relax compressive stress up to a residual value that vanishes \cite{Davidovitch11,Schroll13} with the inverse bendability $\epscap$, see Appendix B.)  Secondly, the radial stress is substantially larger than the stress prior to wetting, $\srr\gg\Tdry$, near the drop. In particular, the tension in the vicinity of the contact line $\To \sim T^\ast = \gamma^{2/3}Y^{1/3}$  (up to logarithmic corrections). Recall also that the stress discontinuity across the contact line is given to leading order in $\tglv$ by $\gamma\cos\theta_Y$; therefore the tension in the sheet is continuous at leading order in $\tglv$, i.e.~$\Ti \approx \gamma^{2/3}Y^{1/3}$ also.  
\\

\underline{\emph{Ultra-weak tension regime}, $ \tauinf < \Rin/\Rout$} (Equivalent to $ \tTdry < \tglv^{2/3}\Rin/\Rout $)

In the perturbative and non-perturbative regimes, the effect of the capillary-induced stress is restricted to a finite zone around the contact line and approaches the uniform, isotropic stress of the pre-wetted state a distance $L \ll \Rout$ from the contact line. (In the perturbative regime, $L$ is proportional to  $\Rin$, while in the non-perturbative regime $ L \sim \Rin/ \tauinf$, see Ref.~  \cite{Schroll13}). 
In contrast, in the ultra-weak tension regime, the drop produces a strong, global effect:  the stress is significantly altered (compared to the pre-wet state) \emph{throughout} the  sheet. If the sheet is clamped, it remains nearly planar, but the radial stress is amplified substantially, reaching a value $\srr(r) \approx T^\ast =\gamma^{2/3}Y^{1/3}$ as $r \to \Rout$. (This is shown as the blue, dash-dotted curves in figs \ref{fig:Profiles}a,b, for the same parameters as in the perturbative regime just discussed but with the addition of a clamped boundary condition at $r=\Rout=10\Rin$.) If the far edge is not clamped, the exerted tensile load is not sufficient to stabilize the planar state, and the sheet wraps around the drop  \cite{Paulsen15}.  
\\

\section{Discussion and critique}

In this section, we point out the limitations of our  FvK-based theoretical approach before discussing  some central assumptions made in previous studies, and  new insights into the validity of these assumptions suggested by our analysis. 

\subsection{Limitations of our FvK-based theoretical framework\label{sec:limitations}}

The FvK equations describe 
solid sheets whose local stress-strain relation is characterized by an isotropic, Hookean response \cite{Landau66}; as such this approach is typically valid only if 
strains are small. 
Since the scales for strain are set by $\glv/Y$ and $\Tdry/Y$, the small strain assumption is compatible with our asymptotic analysis when both extensibility parameters, defined in eqns \eqref{eq:DGSchroll} and \eqref{eq:define-dry-exten}, are small. Furthermore, by using a single planar coordinate system (Monge representation), it is implicitly assumed that the deformed shape represents a small deviation from the planar state, so that, in particular, the slope of the membrane remains small everywhere. 
Since the largest slope of the deformed sheet is given by the inclination  angle $\phi$, which is found to vanish with the extensibility parameters, the small-slope assumption too is self-consistent.       

A further limitation on our results is that they are only valid if the sheet is highly bendable, namely $\epscap \ll 1$; combining this with the
above small strain requirement,  
$\tglv=\glv/Y\ll1$, 
we find that our analysis is limited to sheets whose thickness satisfies \cite{Schroll13}
\begin{equation}
\lm \ll t \ll \Rin^{2/3} \lm^{1/3} \ ,
\label{eqn:asylimit}
\end{equation}
where the length $\lm=\glv/E$ is known in the soft capillarity community as the ``elastocapillary length" \cite{Style17}. Equation \eqref{eqn:asylimit} may alternatively be written:
\begin{equation}
t \ll \lbc \ll \Rin \ ,
\label{eqn:asylimit-2}
\end{equation} where the length $\lbc=(Et^3/\glv)^{1/2}$ (also called an ``elasto-capillary length'' in the geometry-elasticity community \cite{bico04,Duprat16,bico18}), has  recently been termed the ``bendo-capillary length'' for clarity \cite{Style17}.
For $\lbc\ll t\ll\lm$, the response of the sheet may no longer be assumed to be Hookean since then the capillary-induced strain,  
$\glv/Et\gg1$. For drops that are much smaller than 
$\lbc$, the bending energy ($U_{\rm bend}$, see \S2.3 and \S3.1) is no longer small in comparison to the surface energy and strain. Indeed for such small drops, the effect of resistance to bending has a considerable effect on the mechanics \cite{Py07,Neukirch13,Schroll13,Style13,Style17}. (We note that the inequalities above should be slightly modified to account for the fact that the relevant tension scale in the bendo-capillary length $\lbc$ is $\Ti$, not $\glv$; in reality, this represents a small modification of these conditions, as discussed in Appendix B, and so we present the simpler versions here.)

As discussed already in \S1, 
the value of the capillary strain is generally small in the experiments reported to date: with $\glv=72\mathrm{~mN/m}$, glassy films ($E\sim1\mathrm{~GPa}$, $t\gtrsim 100\mathrm{~nm}$) \cite{Schroll13,Toga13,Schulman15} have $\lm\sim0.1\mathrm{~nm}\ll t$ while polymeric films ($E\sim1\mathrm{~MPa}$, $t\gtrsim 1\mathrm{~\mu m}$) \cite{Nadermann13,Schulman15} have $\lm\sim10\mathrm{~nm}\ll t$. The first condition in \eqref{eqn:asylimit} therefore holds in such systems.
The second condition of \eqref{eqn:asylimit} holds provided that the droplets are sufficiently large compared to the 
length $\lbc$, values of which are given in table \ref{table:1}. We see that most sheets (except perhaps the $t=18\mathrm{~\mu m}$ and $t=25\mathrm{~\mu m}$ PDMS sheets \cite{Nadermann13}) satisfy this constraint for droplets of radius $\Rin\gtrsim500\mathrm{~\mu m}$.

We also point out that the upper limit  on the film thickness, $t\ll\Rin^{2/3} \lm^{1/3}$, in \eqref{eqn:asylimit} is strictly different from that of Style \emph{et al.}~\cite{Style17}, who proposed to distinguish between the partial wetting of ``thin" and ``thick" films through the ratio $\glv/Et$ alone, regardless of the droplet's size. Instead, the double inequality in \eqref{eqn:asylimit}, and the analysis in our paper suggest (at least) three qualitatively-distinct types of response for sheets that are free to bend in response to contact with a liquid drop: A non-Hookean response ($t < \lm$); a Hookean, high-bendability response (if both inequalities in \eqref{eqn:asylimit} are satisfied), which has been the subject of this paper (and may be perturbative or non-perturbative depending on the value of $\btauinf$); and a Hookean low-bendability response ($\lbc > \Rin$), which has been the subject of numerous papers \cite{Py07,Neukirch13}.

\subsection{An elasto-capillary probe for pre-tension?\label{sec:pretension}}

We now consider the implications of our results for  understanding the contact between a droplet and  a thin solid sheet. The question of most interest is whether the measured value of $\Ti$ can be used to infer the state of stress in the dry film, $\Tdry$ --- it is this application that has motivated recent studies 
\cite{Nadermann13,Schulman15,Schulman17}. For simplicity, we shall assume a large sheet, \emph{i.e.}~$\Rout/\Rin \gg 1$, and  a given Young's angle: $\theta_Y = \cos^{-1}(\Delta\gamma_s/\glv)$. The assumption of a large sheet ensures that the far-field tension, $\Tinf\approx\Tdry$, as discussed in \S\ref{sec:finitesize}.

Our analysis suggests that the dimensionless membrane inclination angle $\phi\times(Y/\glv\sin\theta_Y)^{1/3}$ gives a simple indication of whether a sheet is in the non-perturbative regime or the perturbative regime. In particular, if 
relationship \eqref{eqn:ruleofthumb} is satisfied,
then we expect that $\Ti\approx\Tinf$ to within $10\%$. Otherwise we expect that the experiments lie in the non-perturbative regime, for which $\Ti\sim T^\ast(\glv,Y) = \glv^{2/3}Y^{1/3}$ (up to a logarithmic dependence on $\Tinf$). Using the criterion in \eqref{eqn:ruleofthumb} suggests that the membrane angles measured by Schulman \emph{et al.} \cite{Schulman15,Schulman17} for  PnBMA sheets and by Nadermann \emph{et al.}~\cite{Nadermann13} for some PDMS sheets are large enough that they lie in the non-perturbative regime. As such, the measured tensions $\Ti,\To$, in the vicinity of the contact line are affected mainly by the capillary-induced stress, $\sim T^\ast(\glv,Y)$, up to a  correcting factor that depends only logarithmically on $\Tinf$. One consequence of this conclusion is that we should expect the measured values of $\Ti,\To$ to be proportional to $t^{1/3}$ (again, up to logarithmic corrections). A plot of the experimentally reported values of $\Ti/\glv $ was shown in fig.~\ref{fig:TiRaw} and, as already discussed, 
is consistent with such a scaling. (However, we emphasize again that for 
the experiments of Schulman \& Dalnoki-Veress~\cite{Schulman15} on SIS membranes, the criterion \eqref{eqn:ruleofthumb} is satisfied, indicating a perturbative response; as such, the $t^{1/3}$ scaling is not expected to be observed for SIS membranes, just as seen in fig.~\ref{fig:TiRaw}.)
\\

Of the experimental data sets discussed in our paper, that for PnBMA sheets \cite{Schulman15} appears to exhibit the strongest non-perturbative response: 
for the sheet thicknesses and Young's  modulus reported in Table 1 (and captions of Fig.~2,3,6), the measured angles, $\phi$, reported in Ref.~\cite{Schulman15} are  large in comparison to $(\glv/Y)^{1/3}$ (see Fig.~6b). Such large values may be observed only if the pre-tension, $\Tpre$, differs greatly from 
the 
value of $\Ti$ in the vicinity of the contact line (see table 1). 
From this perspective, a simultaneous measurement of  these large values of the angle $\phi$, 
corroborated by an independent measurement of  the values of $\Tpre$ reported in Ref.~\cite{Schulman15}, seems possible only if the FvK theory of elastic sheets is not valid for thin PnBMA sheets. According to our discussion in \S5.1, a failure of FvK theory may occur if an amorphous sheet is not a solid phase ({\emph{i.e.}} it has a vanishing shear modulus), or if the material exhibits a strongly non-Hookean response even under the very small strains expected in those experiments, (since here $\glv/Y < 10^{-3}$).

%
It is thus interesting to note that Schulman \& Dalnoki-Veress~\cite{Schulman15} rationalized their measurement of $\Ti$ by comparison with an independent measurement of $\Tdry$ using the indentation of the sheet without any drops present \cite{Schulman15}, 
as well as another estimate of $\Tdry$ that uses the thermal expansion coefficients of Polystyrene membranes \cite{Fortais17}. These results all suggest that $\Ti\approx\Tdry$ (or at least that the discrepancy is relatively small, and not an order of magnitude as we would expect for PnBMA based on  analysis of FvK equations). Yet another indication that the pre-tension, $\Tdry$, is substantially larger than the small values obtained by FvK theory (Table 1) is the reported absence of 
radial wrinkles, which FvK theory predicts for $\btauinf < 0.23$ (see also \S5.3) \footnote{We should note also that a claim made in Ref.~\cite{Nadermann13} and echoed in Ref.~\cite{Schulman15}, on which we will elaborate elsewhere, suggests that a lower bound on the pre-tension in a dry suspended sheet is $2\gamma_{sv}$;  such a putative bound is certainly well above our FvK-based estimates of the pretension in the PnBMA experiments.}. These observations raise the surprising possibility that thin PnBMA sheets (which are believed to be in a glassy state) exhibit a strong non-Hookean response,  invalidating the predictions of FvK theory.     

\subsection{Capillary-induced wrinkles}
As is indicated in fig.~4, for $\btauinf \lesssim0.2297$, our calculations predict that the capillary stress induces hoop compression. Similarly to observations made in the study of a floating sheet by Huang \emph{et al.}~\cite{Huang07} and Toga \emph{et al.}~\cite{Toga13}, such a compression is expected to be relieved through a pattern of radial wrinkles in a ``corona" around the contact line. The extent of this corona has been the object of much study  \cite{Schroll13}. The number of radial wrinkles, $N$, satisfies the scaling law: $N \sim \sqrt{\Rin/\lbc} \sim t^{-3/2} E^{-1/2}$, and is thus expected to be substantially larger in the PnBMA sheets \cite{Schulman15} than in the PDMS sheets \cite{Nadermann13}, while the amplitude of wrinkles is inversely proportional to $N$ and is thus expected to be larger in PDMS sheets than in PnBMA. 
However, we  note that neither Schulman \& Dalnoki-Veress~\cite{Schulman15} nor Nadermann \emph{et al.}~\cite{Nadermann13} reported any observations of such capillary-induced wrinkles. According to our analysis (see Fig.~\ref{fig:Comparison-1}), the absence of capillary-induced wrinkles is plausible for most experiments with PDMS sheets, but certainly not for the experiments with PnBMA sheets, where the corresponding value of $\btauinf$ is much smaller than 0.23. Nevertheless, to verify that hoop compression (and consequently radial wrinkles) may emerge also in a suspended sheet with a clamped boundary, our colleague D. Kumar (UMass Amherst) conducted the demonstration shown in fig.~1(b),(c): an ultra-thin Polystyrene sheet ($t\approx 364 \mathrm{~nm}$, $E\approx3.4\mathrm{~GPa}$) is lifted from a liquid-vapor interface with the aid of a cylindrical cuvette. The two panels fig.~\ref{fig:setup}b,c show a plan view of the suspended sheet (which is believed to be effectively clamped to the cuvette's edge with some unknown pre-tension), before and after a small liquid drop is placed at its center. Kumar's demonstration is clear evidence that capillary-induced wrinkles (and therefore a significant drop-induced perturbation of the pre-stress) are not only a feature of floating sheets, but also emerge in the partial wetting of suspended sheets (for sufficiently small pre-tension).

\subsection{Geometry at the contact line and membrane shape \label{sec:angles}} 

Notwithstanding the crucial distinction between the perturbative and non-perturbative regimes, the underlying idea of extracting the tension in the vicinity of the contact line from the measured contact angles, $\theta$ and $\phi$, remains a sound one. However, there are some important subtleties concerning the measurement of angles in the system, which we now discuss. Specifically we seek to elucidate the following questions: what do we actually mean by the angle $\phi$, and how should it be measured experimentally?

\subsubsection{The angle $\phi$ at different scales}

While the definition of the angle $\phi$ seems  clear, in practice it is not necessarily easy to measure: in our (membrane-theory) approach, the angle $\phi$ is defined to be the angle between the sheet and the horizontal
when viewing the droplet on a scale that is much larger than the bendo-capillary length $\lbc=(B/\glv)^{1/2}$ yet much smaller than the drop's radius, $\Rin$ \footnote{The horizontal scale over which the corner at the contact line is smoothed out by bending stiffness depends on the tension at the contact line, and hence is actually a multiple, $\last$, of $\lbc$ given in \eqref{eq:lbcstar} of Appendix B.}. On such an ``intermediate" scale, both the effect of small-scale curvature (due to bending) and large-scale curvature (due to the spherical shape of the bulged sheet beneath the drop) can be neglected
such that there is a clearly defined angle, $\phi=\phiout$ (see  fig.~\ref{fig:angles}). However, at a scale comparable to $\lbc$, the membrane bends noticeably (since bending stiffness matters at this scale), and the angle between the tangent to the sheet and the horizontal is modified from the angle observed at the outer scale  \emph{i.e.}~$\phiin\neq\phiout$ in the notation of fig.~\ref{fig:angles}.

\begin{figure}
\centering
\includegraphics[width=0.95\columnwidth]{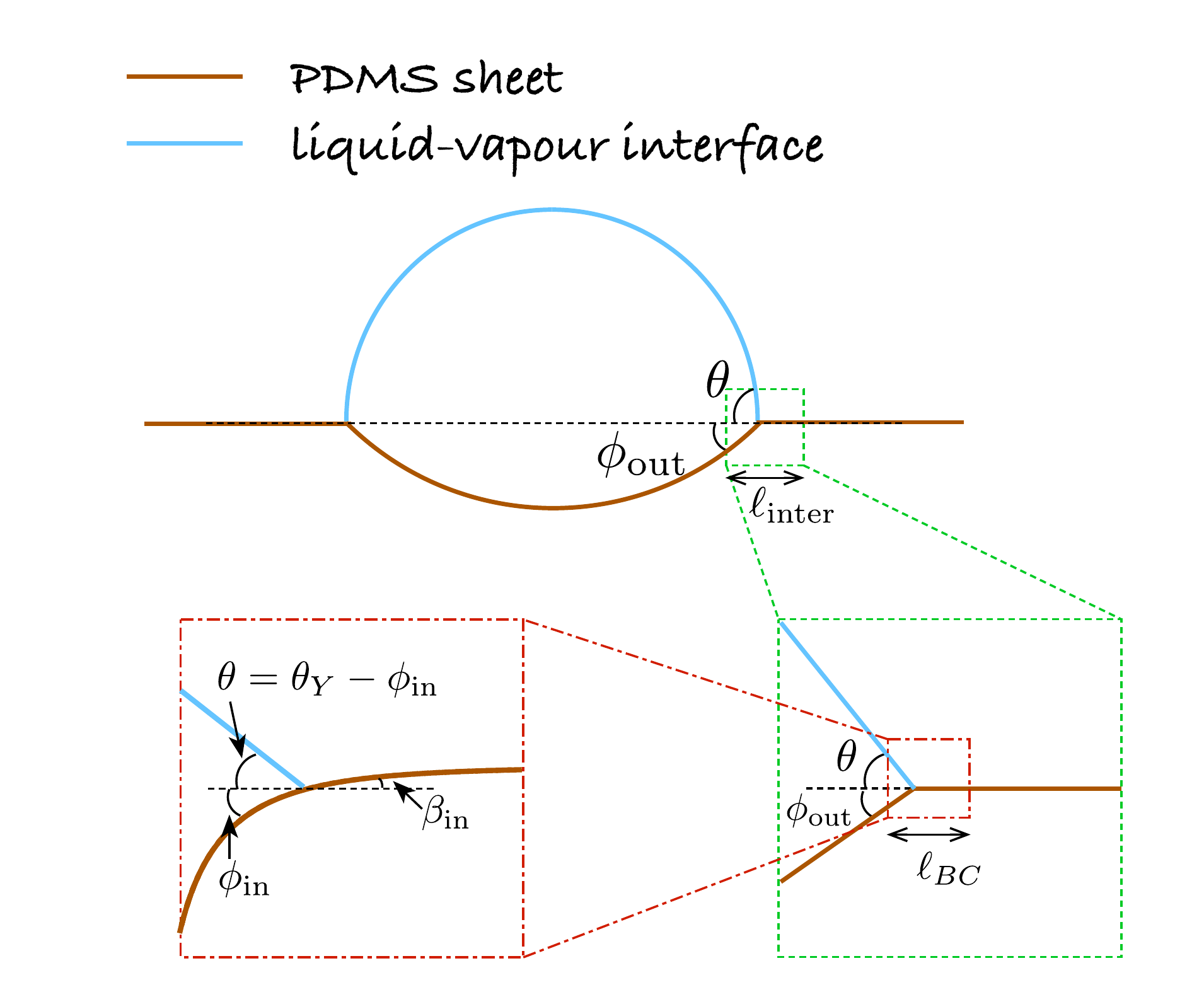}
\caption{Distinction between the different length scales at which the droplet may be observed, and the different angles that result. At an intermediate scale ($\lbc\ll \ell_{\mathrm{inter}} \ll \Rin$), the angle of inclination at the contact line is well-defined, $\phiout$. However, when the contact line is observed at a scale  comparable to the bendo-capillary length, $\lbc\sim(Et^3/\glv)^{1/2}$, the inclination angle will take a different value, $\phiin$.}
\label{fig:angles}
\end{figure}



The distinction between the angles $\phiin$ and $\phiout$ appears to have caused some confusion in the literature. It is our understanding that Nadermann \emph{et al.}~\cite{Nadermann13} measured $\phiin$ and inferred the angle $\theta$ between the liquid-vapor surface and the horizontal from $\theta=\theta_Y-\phiin$ --- a relationship that is expected to be valid when inspecting the contact line at scales $< \lbc$.\cite{Neukirch13} (Note that the symbol ``$\theta$" is used in \cite{Nadermann13} to denote the whole angle between the sheet and the liquid--vapor interface in the vicinity of the contact line (fig.~1d of \cite{Nadermann13}), namely, their "$\theta$" is $\theta  + \phiin$ in our notation.)
In contrast,
Schulman \& Dalnoki-Veress~\cite{Schulman15} measured both $\theta$ and $\phiout$ (in our notation) and observed that $\theta+\phiout\neq \theta_Y$.

To further demonstrate the equivalence of the two approaches we now show that both sets of experiments are consistent with the theoretical prediction of Schroll \emph{et al.}~\cite{Schroll13} who found that
\beq
\frac{\theta_Y-\theta}{\phiout} 
\approx 
\frac{\phiin}{\phiout}
\to \frac{1}{2}
\quad \text{as} \ \tglv\to 0
\label{eqn:GeomClose}
\eeq using a numerical minimization of the total energy in the problem. 
Fig.~\ref{fig:EquiPart} shows that the experimental measurements of both Nadermann \emph{et al.}~\cite{Nadermann13} and Schulman \& Dalnoki-Veress~\cite{Schulman15} are in very good accord with this prediction. In particular, note that the apparent contact angle $\theta+\phiout\neq\theta_Y$ in both sets of experiments. One may also note that any  deviations of the LHS in Eq.~\eqref{eqn:GeomClose} from the value $1/2$ implies deviations of the forces in the vicinity of the contact line from the YLD value ({\emph{i.e.}} $\To - \Ti \neq \gamma\cos\theta_Y$). To see this formally, one needs to consider the higher-order stress terms,  $\bTi^{(2)},\bTo^{(2)}$, in the  $\tglv$-expansion (see Appendix  \ref{sec:compute}), and note that $\theta_Y-\theta \neq \phiout/2 \Rightarrow \bTo^{(2)}-\bTi^{(2)}\neq 0$.  
 Equations~(\ref{eq:bTcont},\ref{eq:bdTDiscont}) in Appendix \ref{sec:compute} shows that this reflects deviations from the YLD contact mechanics (\ref{eq:YLDmechanics-define}), that vanish asymptotically as $\tglv \to 0$. 

\begin{figure}
\centering
\includegraphics[width=0.9\columnwidth]{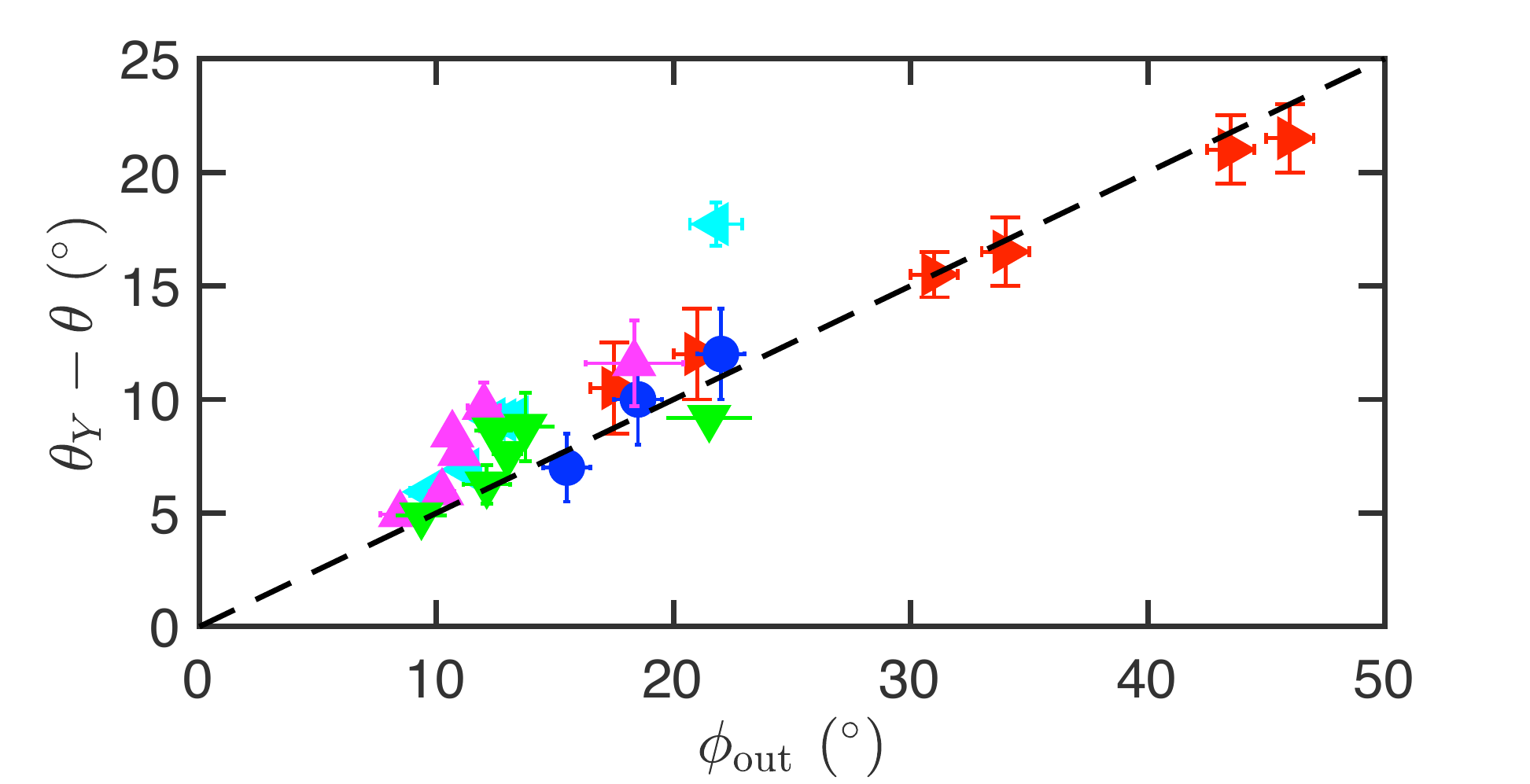}
\caption{Experimental data from Nadermann \emph{et al.}~\cite{Nadermann13} and Schulman \& Dalnoki-Veress~\cite{Schulman15} replotted to show the deviation from the Young angle, $\theta_Y-\theta$ as a function of the inclination of the membrane at the contact line (denoted $\phiout=\phi$ here). The purely geometrical prediction, \eqref{eqn:GeomClose}, is shown by the dashed line. Experimental results are shown for Glycerol drops \cite{Schulman15} on polymeric films of SIS (red right pointing triangles) and glassy films of PnBMA (blue circles); other experimental results are for a variety of droplets on PDMS films \cite{Nadermann13} as follows: De-ionized water (cyan left pointing triangles), Ethylene-Glycol (magenta upward pointing triangles) and DMSO (green downward pointing triangles).}
\label{fig:EquiPart}
\end{figure}

Finally, let us note that while the computation of the tension $\Ti$ ($\sigma_{II}$ in the notation of Nadermann \emph{et al.}~\cite{Nadermann13}) from the angles $\phiin$ and $\phiout$ is consistent, we believe that their computation of $\To$ ($\sigma_{I}$ in their notation), is not precise, since it ignores an inclination in the dry part of the sheet ($\beta_{\mathrm{in}}$ in fig.~\ref{fig:angles}), which, when observing the vicinity of the contact line at scales $<\lbc$, must be taken into consideration. A consistent computation of $\To$ in their approach (from measured 
$\phiout,\phiin$) is simply the YLD law: $\To = \Ti + \glv \cos\theta_Y$ (as was done in \cite{Schulman15}), or, equivalently, $\To = \glv \cos(\theta_Y-\phiin) + \Ti \cos\phiout$, but not
$\To = \glv \cos(\theta_Y-\phiin) + \Ti \cos\phiin$ (Eq.~2 of ref.~\cite{Nadermann13}, expressed in our notations).

\subsubsection{Membrane shapes}

The angle $\phiout$ is typically measured by fitting the membrane shape beneath the drop to a spherical cap \cite{Schulman15}. While this is observed to be a very good description of experimental data, one would expect it only to be strictly valid when the stress in the sheet is very close to being uniform and isotropic, which corresponds to the case of extremely high pre-tension, or the perturbative regime $\btauinf\gg1$. However, our numerical solutions of the full FvK problem (see fig.~\ref{fig:Shapes}) show that, in fact, the shape remains very close to a spherical cap even as $\btauinf$ decreases 
well into the non-perturbative regime, where the stress state is neither uniform nor isotropic. For all but the very smallest values of $\btauinf$, it seems that the spherical approximation is likely to be a good one.

\begin{figure}
\centering
\includegraphics[width=0.8\columnwidth]{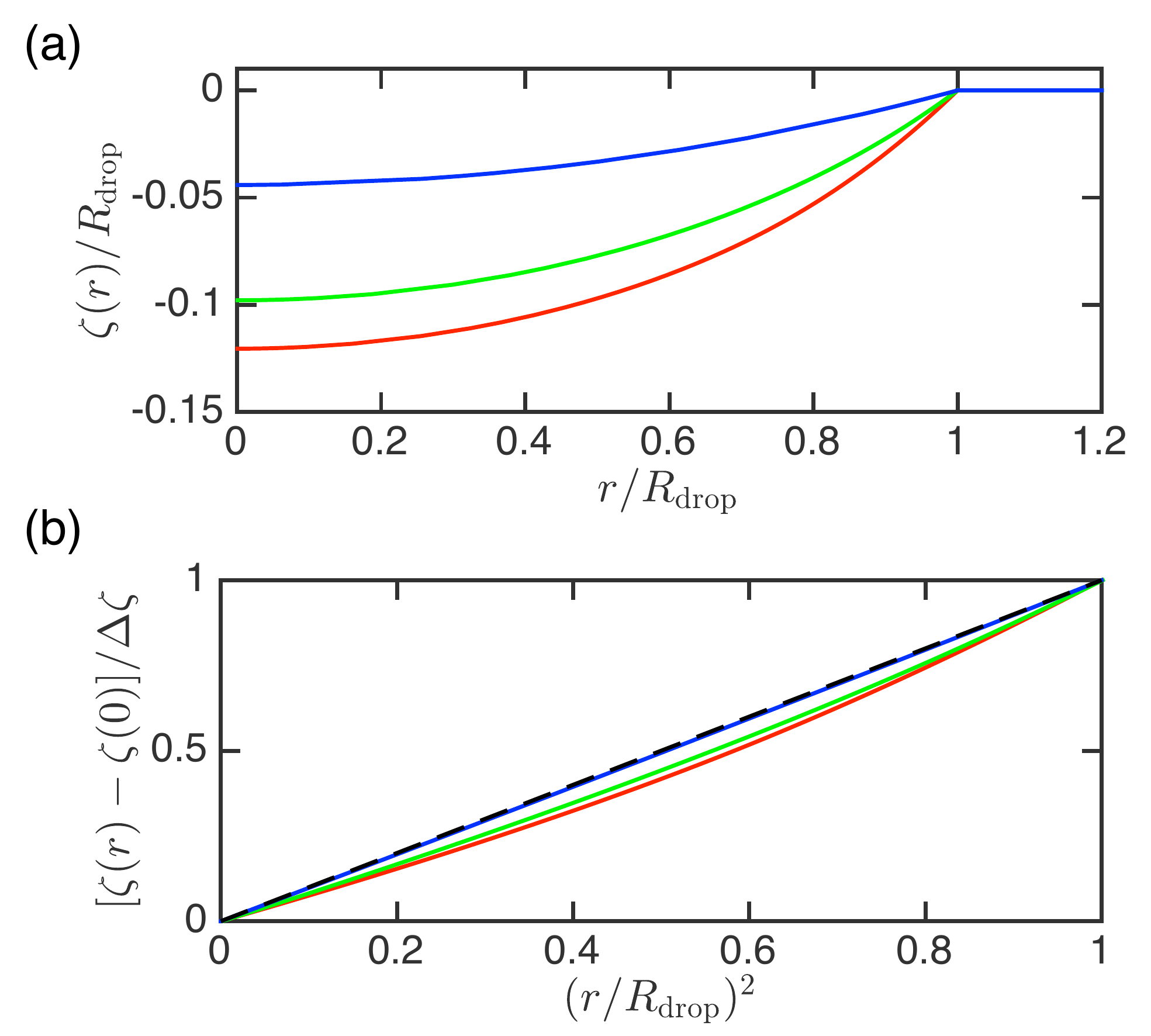}
\caption{The numerically determined membrane shape beneath the drop is very close to a spherical cap. (a) The dimensionless shape for three different values of $\btauinf$: $\btauinf=0.01$ (red curves), $0.1$ (green curves) and $\btauinf=1$ (blue curves). (b) Plotting the rescaled deformation from the centre as a function of $r^2$ highlights how close to spherical caps the membrane shapes in (a) are (a spherical cap corresponds to the line $y=x$, black dashed line). 
Nevertheless, note that  the discrepancy does increase as the pre-tension decreases.}
\label{fig:Shapes}
\end{figure}

\section{Conclusions}

In this paper we have addressed  the partial wetting of a thin solid sheet under tension, following 
Schroll \emph{et al.}~\cite{Schroll13}. 
Our analysis focuses on the limit of {\emph{nearly inextensible}} ($\tglv \ll 1$) yet {\emph{highly bendable}} ($\epscap \ll 1$) sheets. This parameter regime, defined by the inequalities $t \ll \lbc \ll \Rin$ (see \S\ref{sec:limitations}), corresponds to most experiments reported recently on floating \cite{Huang07,Schroll13,Toga13} and suspended sheets \cite{Schulman15,Schulman17,Nadermann13}.   
Our main result, motivated by energetic scaling arguments (\S\ref{sec:qualitative}) and confirmed by a detailed solution of the FvK equations (\S\ref{sec:details},\S\ref{sec:results}), is the presence of two qualitatively distinct types of response: perturbative (in which placing a drop  only affects the uniform, isotropic pre-tension  slightly)  and non-perturbative (in which the capillary-induced stress in the vicinity of the drop is much larger than the pre-tension, and hence depends only weakly on the pre-tension). The borderline between these two regimes is demarcated by the ratio $\tauinf$ (Eq.~\ref{eq:define-tauinf-2}), between the far-field tension, $\Tinf$, and the characteristic capillary-induced stress, $T^\ast \sim \glv^{2/3} Y^{1/3}$.

One counter-intuitive feature of the non-perturbative response in a nearly inextensible sheet ({\emph{i.e.}} $\tglv\ll 1$), is that the capillary-induced stress may dominate a pre-tension $\Tpre$, even if the latter is very large in comparison to $\glv$ (\emph{i.e.}} $\glv \ll \Tpre \ll \glv^{2/3} Y^{1/3}$). A second counter-intuitive feature is that, even though the stress fields are markedly different in each case, the solid-liquid contact in both parameter regimes differs only slightly from the classical YLD law (Eq.~\ref{eq:LaplaceLaw}); the small deviations of the contact angles from YLD contact geometry (\ref{eq:YLDgeometry-define}) are characterized by (distinct) powers of the capillary and dry extensibility parameters, $\tglv, \tTdry$ (see fig.~4 and Eq.~\ref{eqn:PhiAnalytical}).    


A further unusual feature of the partial wetting of a drop of surface tension $\glv$ on a sheet with thickness $t$ is that the capillary-induced stress, $T^\ast \sim \glv (t/\lm)^{1/3}$, \emph{increases} as the elasto-capillary length, $\lm = \glv/E$, decreases. This is the opposite trend to that found in the partial wetting of thick (or non-bendable) solids, where the effect of solid elasticity on liquid contact {\emph{vanishes} as $\lm$ gets sufficiently small. This sharp contrast reflects the nontrivial interplay by which solid geometry and elasticity affect partial wetting phenomena; such an interplay between  geometry and elasticity  may affect solid-liquid interactions in more complex systems, {\emph{e.g.}} a drying colloidal drop that rests on a stretched sheet \cite{Boulogne16}.

Our FvK-based analysis led to a one-to-one relationship between the measured contact angle ($\phi$) and the far-field tension, $\Tinf$, in a large sheet. To facilitate the use of this relationship in experiments, we proposed an expression for this relationship, \eqref{eqn:PhiAnalytical}, that is uniformly valid (up to small errors). We also critically examined
the level of accuracy of previous works \cite{Schulman15,Nadermann13,Fortais17}, which overlooked the possibility of a non-perturbative effect due to capillary-induced stress and instead suggested that the value of any pre-tension in the sheet can be extracted from the contact angle with the aid of local force balance alone. Our analysis of raw data from these works showed that only for a subset of the studied sheets is such a method reliable (in the sense that the inferred value of the pre-tension is correct to within 10\% accuracy); in most instances, the error in the extracted value of pre-tension from local force balance consideration alone ({\emph{i.e.}} without using Eq.(\ref{eqn:PhiAnalytical})) is 50\% and may even be much more. Furthermore, Eq.~(\ref{eqn:PhiAnalytical}) and figs.~\ref{fig:Universal}--\ref{fig:Comparison-1} show that extracting a value of the pre-tension from measurements of the angle $\phi$ is liable to be rather inaccurate if the pre-tension is low enough to lie in the non-perturbative regime (since there $\phi$ depends only very weakly on the pre-tension).   
   





\subsection*{Acknowledgments}

We are grateful to C.~Wong for preliminary calculations in a related problem, and to D.~Kumar for kindly sharing with us his experimental images (fig.~1). We are grateful to A.~Jagota, and to K.~Dalnoki-Veress and R.~Schulman for intense discussions, 
and for generously supplying us with the original data  obtained in their experiments \cite{Nadermann13,Schulman15}. 
We also thank M.~Adda-Bedia, F.~Brau, E.~Cerda, N.~Menon, and R.D.~Schroll, for many discussions, and 
A.~Jagota, K.~Dalnoki-Veress, R.~Schulman, H.A.~Stone, and an anonymous referee for numerous critical comments that helped improve our manuscript. The research
leading to these results has received funding from the European
Research Council under the European Union's Horizon 2020
Programme/ERC Grant Agreement no. 637334 (DV) and NSF-CAREER
Grant No. DMR 11-51780 (BD). We also acknowledge support from the W. M. Keck Foundation. BD benefited from stimulating discussions with participants of the program ``Geometry, elasticity, fluctuations, and order in 2D soft matter'', held in winter 2016 at the Kavli Institute for Theoretical Physics, UCSB.


\begin{appendix}

\section{Details of theoretical approach\label{sec:theory}}

In this section we discuss more completely the full solution of the FvK equations 
that we developed in \S3.

\subsection{Full statement of the problem (unwrinkled version)}


For simplicity, we begin by describing in detail our full solution of the FvK equations in the case where the sheet remains unwrinkled throughout.

\subsubsection{Non-dimensionalization}

The  discussion of \S \ref{sec:perturb} highlights the qualitatively different behaviours that can be observed in the limits $\tauinf\ll1$ and $\tauinf\gg1$. However, for a detailed quantitative analysis of these behaviours, as well as the behaviour for intermediate $\tauinf$, we must first non-dimensionalize the problem \eqref{eqn:FvK1Int}--\eqref{eqn:FvK2-new-out}. To do this, we recall the scaling analysis of \S\ref{sec:qualitative} and in particular the natural tension scale $T^\ast=\glv^{2/3}Y^{1/3}$ introduced in \eqref{eq:T-transit}. We therefore introduce the dimensionless variables:
\begin{equation}
\barpsi =\frac{\psi}{T^\ast\Rin} \ \ ; \ \ 
\barr = \frac{r}{\Rin}
\label{eq:define-nondim-psi}
\end{equation}
 Eq.~(\ref{eqn:FvK2-new-in}) then becomes: 
\begin{gather}
\barr\frac{\upd}{\upd \barr}\left[\frac{1}{\barr}\frac{\upd(\barr \barpsi)}{\upd \barr}\right]= 
-\tfrac{1}{2}\sin^2\theta
\frac{\barr^4}{\barpsi^{2}}
\quad 0\leq \barr\leq 1 \ .
\label{eqn:FvK2ND}
\end{gather} 
Outside the drop, $1\leq\barr\leq \Rout/\Rin$, Eq.~\eqref{eqn:FvK2-new-out} immediately gives the dimensionless version of \eqref{eq:Lameaxi}, namely
\beq
\barpsi=\bTinf\barr+(\bTo-\bTinf)/\barr,
\label{eq:LameaxiND}
\eeq where 
\beq
\bTinf=\Tinf/T^\ast
\eeq and the dimensionless tensile stresses in the immediate vicinity of the contact line are: 
\begin{equation}
\bTi = \frac{\srr(r = \Rin^-)}{T^\ast},\quad
\bTo =\frac{\srr(r = \Rin^+)}{T^\ast}.
\label{eq:define-nondim-stress}
\end{equation}

\subsubsection{Local boundary conditions\label{sec:BCs}}

The equation for the stress in the sheet beneath the droplet, (\ref{eqn:FvK2ND}), is a second order differential equation, requiring two BCs. One BC is that  the  horizontal displacement vanishes at the centre of the drop, {\emph{i.e.}}~$u_r(r\to0)=0$ or, in terms of the stress potential,
\begin{equation}
\lim_{\barr\to0}[\barr\barpsi'-\nu\barpsi]=0.
\label{eq:BCcenter} \end{equation}
The second BC is that the tension at the contact line (approached from within the drop) is $\Ti$, {\emph{i.e.}}
\begin{equation}
\bTi= \barpsi(1^-)\ ,  
\label{eq:BCRd}
\end{equation}
defining $\bTi$ as a first unknown.
We denote the solution of Eq.~(\ref{eqn:FvK2ND}) subject to the two BCs \eqref{eq:BCcenter} and \eqref{eq:BCRd} by $\barpsiin(\barr;\bTi,\theta)$. The function $\barpsiin(\barr;\bTi,\theta)$ must be determined numerically \cite{Vella10,King12} as we will describe below.

For a given  $\tauinf$, the stress within the dry part of the sheet ($\barr>1$) is given, in terms of a single unknown $\bTo$, by the dimensionless Lam\'{e} solution, \eqref{eq:LameaxiND}. To contrast with the solution for the wetted portion of the sheet, we denote this by the function $\barpsiout(\barr;\bTo,\tauinf)$ 
(We emphasize that the stress associated with $\barpsiout$ remains tensile (\emph{i.e.}~positive) everywhere
provided that $\bTo<2\tauinf$; otherwise this solution must be replaced by an 
analogous result, which incorporates the effect of wrinkles, 
see eqn \eqref{eq:LameFT} below.)

The stress potential beneath the drop, $\barpsiin$, depends on two unknowns, $\bTi$ and $\theta$; together with the angle of the sheet at the contact line, $\phi$, and $\bTo$, we have four unknowns in total. The two (normalized) tensions $\bTi,\bTo$, and the two angles $\theta,\phi$, are related by  force balance equations at the contact line:
\begin{gather}
\bTo = \tglv^{1/3}\cos\theta + \bTi\cos\phi \label{eq:H-balance} \ , \\
\bTi\sin\phi = \tglv^{1/3}\sin\theta  \ , 
\label{eq:V-balance} 
\end{gather} respectively, where we recall the definition $\tglv=\glv/Y$, \eqref{eq:DGSchroll}. (Note that Eq.~(\ref{eq:V-balance}) is merely an evaluation of (\ref{eqn:FvK1Int}) as $r \to\Rin^-$.)

A further equation connecting the four unknowns is the continuity of radial displacement at the contact line, 
\beq \rmurin (\barr;\bTi,\theta)|_{\barr \to 1^{-}} =  \rmurout (\barr;\bTo)|_{\barr \to 1^{+}} \ ,  
\label{eq:radialdisp-cont1} 
\end{equation}
where $\rmurin,\rmurout$ are functions that determine the radial displacement, respectively, within the wet and dry zones of the sheet. 
These functions may be determined in terms of $\barpsiin$ and $\barpsiout$ since (in axisymmetry) the radial displacement $u_r=r\epsilon_{\theta\theta}$ and the hoop strain $\epsilon_{\theta\theta}=(\sqq-\nu\srr)/Y$ (from Hooke's law). Hence, relating stress to the potentials $\barpsiin$ and $\barpsiout$, we may write \eqref{eq:radialdisp-cont1} as
\beq
\left.\frac{\upd\barpsiin}{\upd r}\right|_{\barr=1}-\nu\barpsiin(1;\bTi,\theta)=
2\bTinf-(1+\nu)\bTo
\label{eq:radialdisp-cont} 
\eeq
\\

Equations \eqref{eq:H-balance}, \eqref{eq:V-balance} and \eqref{eq:radialdisp-cont} 
comprise three equations for the four unknowns $\bTi,\bTo$, $\theta$, and $\phi$. We therefore require another condition to close the problem; we address this missing equation next.

\subsubsection{Non-local effects at the contact line\label{sec:contactline}} 

Since the YLD angle $\theta_Y$ does not appear explicitly in any of the equations \eqref{eq:H-balance}--\eqref{eq:radialdisp-cont},  it is tempting to propose a fourth equation by analogy with the classical YLD contact limit (\ref{eq:YLDcontact-define}). 
Two proposals for this missing link  are: (i) that $\theta=\theta_Y$ (Ref. \cite{Vella10}), and (ii) $\bTo - \bTi = \tglv\cos\theta_Y$ (Ref.\cite{Schulman15} and an analogous assumption in Ref.~\cite{Nadermann13}). We  discuss the practical value of these proposals in \S\ref{sec:angles}; however, let us note that the mere search for a simple rule overlooks an important conceptual aspect of the partial wetting problem. As was pointed out by Olives \cite{Olives93}, a complete characterization of the contact requires a {\emph{global minimization of the energy}}, namely,  $\Ustrain + \Ubend + \Usurf$ (see \S2.3), and cannot be determined from considerations of local force balance alone. 

Olives's insight was taken up by Schroll \emph{et al.}~\cite{Schroll13}, who used the three equations \eqref{eq:H-balance}, \eqref{eq:V-balance} and \eqref{eq:radialdisp-cont} to eliminate three of the four unknowns and then computed the total energy, $\Ustrain + \Usurf$ (exploiting the negligibility of $\Ubend$ in the high bendability regime), as a function of the single remaining variable, which was taken for convenience to be the angle $\phi$. They then minimized this energy as a function of the angle $\phi$ to obtain their final solution. 

To avoid the need for a numerical minimization of the total energy, and to shed some light on the underlying physics, we employ here an analytical approach to the problem, by expanding around the YLD limit~(\ref{eq:YLDcontact-define}). Namely, we assume that the contact angles $\phi$ and $\theta$, are described by a power series: \begin{eqnarray}
\phi &=& \phi_1 \tglv^{\beta} + 
\cdots 
\label{eq:expand-phi}  \\
\theta &=& \theta_Y + \dtheta_1 \tglv^{\beta} +\cdots
%
\label{eq:expand-dtheta} 
\end{eqnarray}   
such that the YLD contact geometry, $\phi \to 0, \theta\to \theta_Y$ is approached asymptotically as $\tglv \to 0$, with some (as yet undetermined) exponent $\beta$ and coefficients $\phi_i$ and $\dtheta_i$. The existence of such an expansion is motivated by the qualitative discussion in \S2, and by experimental and numerical results \cite{Schroll13} (see Appendix A.3). 

In the current paper, we will focus on the leading order behaviour of this expansion, which provides the numerical accuracy necessary for a reliable prediction of the far-field tension, $\Tinf$, 
from the measured angles in the parameter regime studied in  
recent experiments on suspended sheets \cite{Schulman15,Nadermann13}. As we will discuss below, such a leading-order analysis is insufficient to identify the  deviations of the stress jump at the contact line, $\To - \Ti$ from the 
YLD contact mechanics (\ref{eq:YLDmechanics-define}). 
Such corrections can emerge only as higher powers of the parameter $\tglv$ --- a conceptually important issue that will be addressed in a future publication, where we will proceed to compute higher orders in this expansion.       

\subsubsection{Computational scheme}\label{sec:compute}

To translate the solution of \eqref{eqn:FvK2ND} with Eqs.~\eqref{eq:H-balance}--\eqref{eq:expand-dtheta}
into a computational scheme, we extend the expansion in powers of $\tglv$ used in Eqs.~(\ref{eq:expand-phi},\ref{eq:expand-dtheta}), to express the stresses, $\bTi, \bTo$ also:
\beq
\bTi=\bTi^{(0)}+\tglv^{\beta}\bTi^{(1)}+..
\label{eqn:bTiExpand}
\eeq and
\beq
\bTo=\bTo^{(0)}+\tglv^{\beta}\bTo^{(1)}+...
\label{eqn:bToExpand}
\eeq

Substituting the expressions \eqref{eqn:bTiExpand} and \eqref{eqn:bToExpand} into the force balances at the contact line, Eqs.~\eqref{eq:H-balance} and \eqref{eq:V-balance}, we consider terms up to $O(\tglv^\beta)$  to find that: 
\beq
\bTo^{(0)}+\tglv^\beta\bTo^{(1)}= \tglv^{1/3}\cos\theta_Y+\bTi^{(0)}+\tglv^\beta\bTi^{(1)}+O(\tglv^{2\beta})  \label{eq:expand-bdT} 
\eeq and
\beq
\tglv^\beta\bTi^{(0)} \phi_1=\tglv^{1/3} \sin\theta_Y+O(\tglv^{\beta+1/3}).
\label{eq:expand-bTi} 
\eeq

Inspection of \eqref{eq:expand-bdT} and \eqref{eq:expand-bTi} implies that $\beta=1/3$, and hence that: 
\beq
\bTo^{(0)}=\bTi^{(0)},
 \label{eq:bTcont} 
\eeq
\beq
\bTo^{(1)}=\bTi^{(1)}+\cos\theta_Y
 \label{eq:bdTDiscont} 
\eeq and
\beq
\bTi^{(0)} \phi_1= \sin\theta_Y.
\label{eq:phi1explicit} 
\eeq

To proceed, we let
\beq
\barpsiin(\barr)=\barpsiin^{(0)}(\barr)+\tglv^{1/3}\barpsiin^{(1)}(\barr)+O(\tglv^{2/3})
\label{eqn:psiinExp}
\eeq with a similar expansion for $\barpsiout(\barr)$. The leading-order of \eqref{eqn:FvK2ND} is easily seen to be
\beq
\barr\frac{\upd}{\upd \barr}\left[\frac{1}{\barr}\frac{\upd\bigl(\barr \barpsiin^{(0)}\bigr)}{\upd \barr}\right]= 
-\tfrac{1}{2}\sin^2\theta_Y
\frac{\barr^4}{\bigl(\barpsiin^{(0)}\bigr)^{2}}
\quad 0\leq \barr\leq 1 \ .
\label{eqn:FvKpsiin0}
\eeq

Substituting 
\eqref{eqn:psiinExp} into Eq.~(\ref{eq:radialdisp-cont}), and 
making use of \eqref{eq:bTcont} we obtain a boundary condition for \eqref{eqn:FvKpsiin0}: 
\beq
\left.\frac{\upd\barpsiin^{(0)}}{\upd\barr}\right|_{\barr=1}+\barpsiin^{(0)}(1)=2\bTinf.
\label{eq:radialdisp-cont-1} 
\end{equation} 
Together with the leading-order term of \eqref{eq:BCcenter}, \emph{i.e.}
\beq
\lim_{\barr\to0}\left[\barr\frac{\upd\barpsiin^{(0)}}{\upd \barr}-\nu\barpsiin^{(0)}\right]=0,
\label{eq:BCcenterFirstOrder}
\eeq we then have two boundary conditions for the second-order ODE \eqref{eqn:FvKpsiin0}. At leading order in $\tglv$ the problem in $0\leq\barr\leq1$ is then completely specified and can be solved for given $\tauinf$ and $\theta_Y$. Combining this solution with Eq.~(\ref{eq:bTcont}) and (\ref{eq:phi1explicit}), 
the contact is then characterized 
at leading order in $\tglv$. (Note that the leading-order behaviour of the angular deflection of the membrane at the contact line, $\phi=\tglv^{1/3}\phi$ is slaved to the leading-order stress at the contact line through \eqref{eq:phi1explicit}.)


A few comments are in order:
\\

$\bullet$ Note that the deviation $\dtheta$ of the upper angle from $\theta_Y$ {\emph{is not}} determined by this leading order calculation --- its evaluation  requires carrying out the expansion of Eq.~(\ref{eq:radialdisp-cont}) to higher orders in $\tglv$. Similarly, the deviation of  $\To-\Ti$ from its YLD value ($\glv\cos\theta_Y$), occurs at $O(\tglv^{2/3})$ and hence is beyond  the leading order calculation presented here.
This observation (and more generally, the structure of the leading order equations (\ref{eq:bTcont}-\ref{eq:phi1explicit})), mirrors our qualitative analysis in \S2, which required only the assumption  $\dtheta \to 0$ as $\tglv \to 0$ to find the leading terms in $\phi$ and $\Ti$. 

$\bullet$ Recall that the parameter $\tauinf$ is assumed to take a fixed value in the expansion \eqref{eq:expand-phi}--\eqref{eqn:bToExpand}, and thus the solution of the leading order equations yields the angle 
$\phi(\tauinf,\glv) \approx \phi_1(\tauinf) \tglv^{1/3}$. The numerically-determined function $\phi_1(\tauinf)$, reported  in \S\ref{sec:results}, does exhibit the scaling rules that were anticipated by the qualitative analysis of \S2, {\emph{i.e.}}~eqns \eqref{eq:approachYoung} and \eqref{eq:approachYoung-2} are recovered, up to some logarithmic corrections, in the expected parameter regimes ($\tauinf \ll 1$ and $\tauinf \gg 1$, respectively).

\subsection{Membrane theory versus tension field theory}


In general, the problem of a drop sitting on an elastic sheet for $\tglv \ll 1$, can be divided into two regimes depending on the value of the dimensionless parameter $\tauinf$, which in turn determines the nature of the ``membrane theory" solution (\emph{i.e.} the solution of the FvK equations (\ref{eqn:FvK1},\ref{eqn:FvK2}) neglecting the explicit, high-order effect of bending terms). 
These two regimes are:

(a) A parameter regime in which membrane theory yields a stable solution, {\emph{i.e.}} both the radial and hoop stresses of the membrane solution are purely tensile ({\emph{i.e.}} positive) everywhere. This solution is obtained by solving \eqref{eqn:FvKpsiin0} subject to \eqref{eq:radialdisp-cont-1} and \eqref{eq:BCcenterFirstOrder}.

(b) A parameter regime in which membrane theory, {\emph{i.e.}} the solution of Eqs.~(\ref{eqn:FvKpsiin0}-\ref{eq:BCcenterFirstOrder}), predicts a state with a negative hoop stress, $\sqq(r)<0$, in an annular zone that includes the contact line; such a solution is unstable to the formation of radial wrinkles.  

In \S \ref{App:Membrane} we describe the membrane theory solution, while in \S \ref{App:Wrink} we discuss the ``tension field'' solution that characterizes the wrinkled state in the limit of high bendability, $\epscap \ll 1$ (Eq.~\ref{eq:define-bendability}).

\subsubsection{Axisymmetric deformations (membrane theory)\label{App:Membrane}}

We consider Eqs.~
\eqref{eqn:FvK2ND} and \eqref{eq:LameaxiND} in their respective intervals, 
$0\leq 
\barr\leq1$ and 
$1\leq\barr\leq\Rout/\Rin$. In the first region (the wet tensile region beneath the drop) the leading--order problem \eqref{eqn:FvKpsiin0}--\eqref{eq:BCcenterFirstOrder} may be rescaled by letting 
$\tpsiin=\barpsiin/\sin^{2/3}\theta_Y$ to give 
\beq
\barr\frac{\upd}{\upd\barr}\left[\frac{1}{\barr}\frac{\upd}{\upd \barr}(\barr\tpsiin^{(0)})\right]=-\tfrac{1}{2}\frac{\barr^4}{\bigl[\tpsiin^{(0)}\bigr]^2}
\label{eqn:LeadOrdAxisym}
\eeq subject to the boundary conditions
\begin{equation}
\lim_{\barr\to0}\bigl[\barr \frac{\upd\tpsiin^{(0)}}{\upd \barr}-\nu\tpsiin^{(0)}\bigr]=0,\quad \left.\frac{\upd\tpsiin^{(0)}}{\upd\barr}\right|_{\barr=1}+\tpsiin^{(0)}(1)=2\btauinf
\label{eq:AxisymBCS}
\end{equation} with $\btauinf=\tauinf/\sin^{2/3}\theta_Y$, as defined in \eqref{eq:define-tauinf-2}. 

The problem \eqref{eqn:LeadOrdAxisym} subject to \eqref{eq:AxisymBCS} contains only the single parameter $\btauinf$ and may readily be solved numerically using, for example, the MATLAB routine \texttt{bvp4c}. 
Once this numerical solution in the wetted region $0\leq\barr\leq1$ has been determined for a given value of $\btauinf$, the value of $\bTo=\tpsiin^{(0)}(\barr=1)\sin^{2/3}\theta_Y$ can be determined for a given $\theta_Y$, and the solution in the dry region $1\leq\barr\leq\Rout/\Rin$ read off from \eqref{eq:LameaxiND}. 
We find numerically that the stress remains tensile everywhere, $\srr,\sqq>0$, provided that $\btauinf\gtrsim0.2297$. This defines the parameter regime for which the solution provided by membrane theory is stable.

\subsubsection{Wrinkled state (tension field theory)\label{App:Wrink}}

For $\btauinf\lesssim0.2297$,  membrane theory yields $\sqq<0$ somewhere in the sheet; in reality such a compression would be relaxed, $\sqq \approx 0$ in some region $\Li<r<\Lo$ (where the $\approx$ sign indicates terms that vanish as the bendability, defined in \eqref{eq:define-bendability}, $\epsilon \to 0$). Such an asymptotically compression-free solution to Eqs.~(\ref{eqn:FvK2-new-in},\ref{eqn:FvK2-new-out}), is described by tension field theory \cite{Davidovitch11,King12,Schroll13}.

To obtain the tension field theory solution, the sheet must be divided into four spatial regions that are each treated differently: a wet tensile region ($0\leq r\leq \Li$), a wet wrinkled region ($\Li\leq r\leq \Rin$), a dry wrinkled region ($\Rin\leq r\leq\Lo$) and a dry tensile region ($\Lo\leq r\leq \Rout$). These regions must be joined together by appropriate matching conditions where they meet. The equations that are relevant in each region (at leading order in $\tglv$), together with the appropriate matching conditions are the subject of this subsection.

Relaxing the compressive stress associated with wrinkling, \emph{i.e.} setting $\sqq\approx0$, we find that $\srr=C/r$ throughout the wrinkled region $\Li\leq r\leq\Lo$. (In particular, the same constant $C$ holds for $\Li\lesssim r\lesssim \Rin$ and $\Rin\lesssim r\lesssim \Lo$ by the condition $\bTi^{(0)}=\bTo^{(0)}$.) We may therefore write that the stress potentials are
\begin{equation}
\barpsiin(\barr) = \begin{cases}
\bar{\psi}_t(\barr;\Li,\btauinf), \quad 0\leq \barr< \Li/\Rin\\
\bTo, \quad\quad\quad \Li/\Rin<\barr<1   \ ,\end{cases}\label{eq:InnerFT} 
\end{equation} and
\begin{equation}
\barpsiout(r) = \begin{cases}
\bTo, \quad \quad\quad \quad \quad1<\barr<\Lo/\Rin \\
\bTinf\left( \barr+\frac{\Lo^2/\Rin^2}{\barr}\right), \quad \Lo/\Rin< \barr< \Rout/\Rin  \ .\end{cases}\label{eq:LameFT} 
\end{equation}

The stress potential in the inner tensile region,
$\bar{\psi}_t(\barr;\Li,\btauinf)$ that appears in \eqref{eq:InnerFT} solves \eqref{eqn:FvKpsiin0} but with boundary conditions 
\begin{equation}
\lim_{\barr\to0}\bigl[\barr \frac{\upd\bar{\psi}_t}{\upd \barr}-\nu\bar{\psi}_t\bigr]=0 \ ,
\left.\frac{\upd\bar{\psi}_t}{\upd\barr}\right|_{\barr=\Li/\Rin}\!\!\!\!\!\!=0 \  ,  
\left.{\bar{\psi}_t}\right|_{\barr=\Li/\Rin} = \bTo \, 
\label{eq:UniversalBCs}
\end{equation} 
where the last two express continuity of stress and displacement fields at $r=L_I$ \cite{Davidovitch11}. 

Matching the radial stress at $\barr=\Lo/\Rin$, we find that
\begin{equation}
\bTo=2\bTinf\Lo/\Rin.
\label{eqn:ToCondFT}
\end{equation} (Note that continuity of the hoop stress at $\barr=\Lo/\Rin$ was used already in writing down the specific Lam\'e form of \eqref{eq:LameFT}.) Matching the stresses at $\barr=\Li/\Rin$ we find that
\begin{equation}
\frac{\btauinf^3\Lo^3\Rin^2}{\Li^5}=k,
\label{eqn:FTking}
\end{equation} where $k\approx0.012$ is a constant that emerges from solving the inner tensile problem numerically \cite{King12}.

For a given $\btauinf<0.2297$, we therefore have three unknowns ($\To$, $\Li$ and $\Lo$) with two equations relating them, \eqref{eqn:ToCondFT} and \eqref{eqn:FTking}. A final condition is obtained by requiring that $u_r(\Li^+)=u_r(\Lo^-)$.  (This result follows from continuity of the radial displacement across the edges of the wrinkled zone, \emph{i.e.} $u_r(\Li^-)=u_r(\Li^+)$ and $u_r(\Lo^-)=u_r(\Lo^+)$, combined with the vanishing of the hoop stress within the wrinkled zone, which gives that $u_r(\Li^-)=C/Y=u_r(\Li^+)$.) We therefore have that
\begin{equation}
0=[u_r]_{\Li}^{\Lo}=\int_{\Li}^{\Lo}\frac{\partial u}{\partial r}~\upd r=\int_{\Li}^{\Lo}\frac{\srr}{Y}-\tfrac{1}{2}\left(\frac{\upd\zeta}{\upd r}\right)^2~\upd r,
\end{equation} which, upon using $\srr=\To\Rin/r$ and the membrane shape \cite{King12,Schroll13}
\begin{equation}
\zeta(r)=\begin{cases}
\frac{\glv\sin\theta_Y\Rin^2}{6\Tinf\Lo}(\barr^3-1),\quad \Li/\Rin\leq \barr\leq 1\nonumber\\
0,\quad \quad\quad\quad \quad 1\leq \barr\leq \Lo/\Rin,
\end{cases}
\end{equation} gives a final (closing) relationship
\begin{equation}
\btauinf^3\frac{\Lo^3}{\Rin^3}\log\frac{\Lo}{\Li}=\frac{1}{80}(1-\Li^5/\Rin^5).
\label{eqn:FTtrans}
\end{equation}

 This transcendental equation provides a closing equation, and the  system of equations \eqref{eqn:ToCondFT}, \eqref{eqn:FTking} and \eqref{eqn:FTtrans} can readily be solved numerically to give, for example, $\bTo(\tauinf)$.
 
To make further progress, it is useful to note that \eqref{eqn:FTking} can be used to eliminate $\Lo$ in favour of $\Li$:
 \begin{equation}
 \frac{80k}{3}\log\left(\frac{k\Li^2}{\btauinf^3\Rin^2}\right)=\frac{\Rin^5}{\Li^5}-1.
 \end{equation} This expression can readily be inverted to give the value of $\btauinf$ that would lead to a given wrinkle inner position $\Li/\Rin$
 \begin{equation}
 \btauinf={k}^{1/3}\left(\Li/\Rin\right)^{2/3}\exp\left[\frac{1}{80k}(1-\Rin^5/\Li^5)\right].
 \label{eqn:wrinksExact}
 \end{equation}
 
 The expression in \eqref{eqn:wrinksExact} is used to calculate the behaviour plotted in figures in the main text. However, of particular interest is the limit $\btauinf\to0$. In this limit, we further expect that $\Li/\Rin\ll1$ and hence find that
 \begin{equation}
 \frac{\Li}{\Rin}\approx\left[80k\left(\tfrac{1}{3}\log k-\log\btauinf\right)\right]^{-1/5},
 \end{equation}  which can then be combined with \eqref{eqn:ToCondFT} and \eqref{eqn:FTking} to give
 \begin{equation}
 \Ti\approx \gamma^{2/3}Y^{1/3}\sin^{2/3}\theta_Y\left(C-10\log\btauinf\right)^{-1/3}
 \label{eqn:AppATin}
 \end{equation} where $C=(8k)^{-1}+(10\log k)/3\approx-4.394$. 
Although this asymptotic result is formally only valid for $\btauinf\ll1$, we find that in fact this expression is accurate to within $3\%$ of the true, numerically computed value, for all $\btauinf\lesssim0.2297$. We therefore suggest that this should be used in all wrinkled cases, $\btauinf\lesssim0.2297$; 
Eq.~(\ref{eqn:AppATin}) motivates \eqref{eqn:PhiAnalytical}.  

\subsection{Validity of asymptotic expansion}

Being the leading-order term in an expansion, the solution of \eqref{eqn:LeadOrdAxisym}--\eqref{eq:AxisymBCS}
(and the equivalent in the wrinkled case) is expected to be rather accurate at sufficiently small values of $\tglv$. A quantitative estimate for the accuracy is provided by comparing the value of $\phi \approx \phi_1(\tauinf)\tglv^{1/3}$ obtained from the leading-order solution, with the full energy minimization analysis of Schroll \emph{et al.}~\cite{Schroll13}. Plotting the two solutions as a function of $\tglv$, see fig.~\ref{fig:CompareLeadOrd}, we find that  our leading-order approach provides very good accuracy provided that $\tglv\lesssim10^{-2}$, which is the parameter regime of most of the experiments that we address in \S\ref{sec:results}.    

\begin{figure}
\centering
\includegraphics[width=0.9\columnwidth]{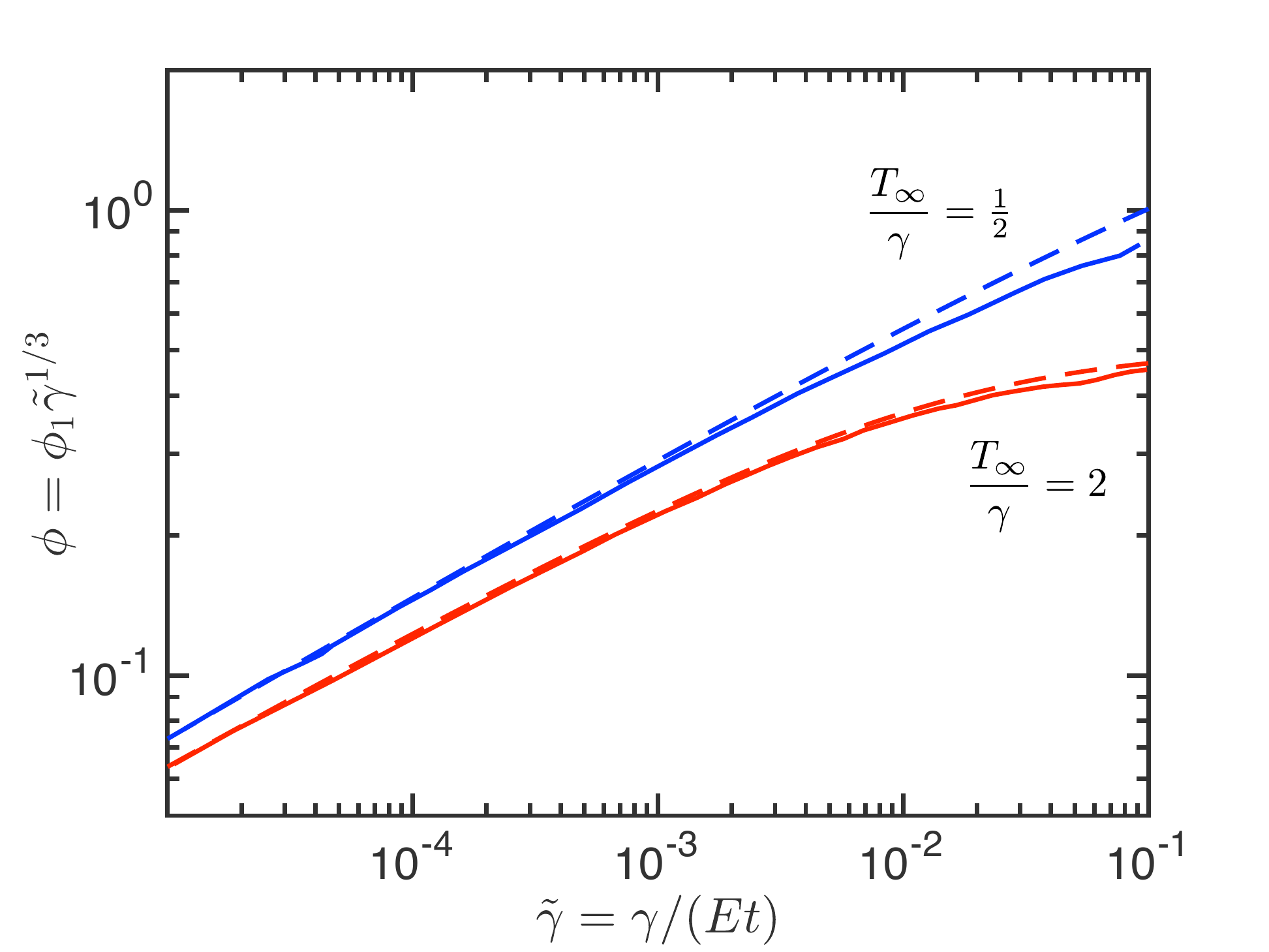}
\caption{The dependence of the angle $\phi$ on the parameter $\tglv$ for two different values of the pre-tension $\tauinf=1/2$ (red curves) and $\tauinf=2$ (blue curves) for $\theta_Y=\pi/2$. The full numerical results of Schroll \emph{et al.}~\cite{Schroll13} (solid curves) are well approximated by the leading order of our perturbation expansion (dashed curves), provided that $\tglv\lesssim10^{-2}$.}
\label{fig:CompareLeadOrd}
\end{figure}

\section{Neglecting bending rigidity\label{sec:bending}}
Our analysis and results do not include any explicit dependence on the bending modulus, $B$, since our study addresses the limit of high capillary bendability, $\epscap \ll 1$ (Eq.~\ref{eq:define-bendability}). Here 
we briefly explain the perturbative (yet singular) effect of bending rigidity on the mechanics in this parameter regime (the interested reader is referred to earlier work \cite{Davidovitch11,King12,Schroll13,
Vella17,Taffetani17} for an expanded discussion of the effect of  bending rigidity in this class of elasto-capillary problems). 
\\

{\emph{Boundary layer:}} As the schematic drawing fig.~\ref{fig:angles} shows, the sheet appears to have a sharp corner at the contact line when observed at scales larger than some length scale, $\last$. However, in reality this corner is smoothed by the small (but finite) bending stiffness of the sheet --- an effect that is visible at scales comparable to $\last$. The scale $\last$ can be determined by balancing the term that would represent bending stiffness in the vertical force balance \eqref{eqn:FvK1}, $B\,\upd^4\zeta/\upd r^4$, with the tensile term in the vicinity of the contact line, $\sim \Ti\,\upd^2\zeta/\upd r^2$. The width, $\last$, of the ``boundary layer" that results from this balance is:  
\begin{equation} 
\last = (B/\Ti)^{1/2} \sim \lbc\cdot \left(\frac{\glv}{\Ti}\right)^{1/2} \sim \frac{\lbc}{\left[
\max\{\tfrac{\Tinf}{\glv},\left(\tfrac{Y}{\glv}\right)^{1/3} \}\right]^{1/2}} \ . 
\label{eq:lbcstar}
\end{equation} 

Our analysis is valid under the assumption that the scales are suitably well-separated, \emph{i.e.}~$t\ll \last\ll \Rin$. One may easily verify that if the two inequalities in \eqref{eqn:asylimit-2} are satisfied then the above inequalities are also satisfied.

Within a typical horizontal distance, $\last$, of the contact line, the  bending energy (and the consequent force) is non-negligible, implying substantial deviations of the  shape, $\zeta(r)$, and stress potential, $\psi(r)$, from the sharp-corner solution described by \S3 and \S4 (which arose from the minimization of the surface and strain energies, but neglecting bending energy). The energetic cost of bending, $\Delta \Ubend$, can be estimated as $\Delta \Ubend \sim 2\pi\Rin \last \cdot B/\last^2 \sim \epscap^{1/2}\, \Delta \Usurf\ll \Delta \Usurf$, justifying the energetic hierarchy, Eq.~(\ref{eq:UHeirarchy}) invoked in \S2.4.

Furthermore, we expect the effect of the boundary layer on the stress field  or shape to be a perturbation of the solutions shown in fig.~\ref{fig:Profiles}  and fig.~\ref{fig:Shapes}, entering at $O(\last/\Rin) \sim O(\epscap^{1/2})$; as such, this perturbation can be safely ignored (see also page 1 of supplementary information in Ref.~\cite{Nadermann13}).  In a purely tensile (unwrinkled) state ({\emph{i.e.}} where the axisymmetric, membrane theory solution is stable), this boundary-layer effect is the primary contribution of bending energy, hence the above argument implies that one can safely ignore any effect, explicit or implicit, of the bending rigidity on the mechanics.  
\\

{\emph{Wrinkled state:}} 
If the stress obtained by membrane theory 
has a compressive zone 
then the sheet is unstable to the formation of wrinkles, which act to relax compression (see Appendix A). Naively, one may assume that the bending rigidity, which clearly governs the wavelength of wrinkles, affects also the stress field. However, the basic premise of tension field theory \cite{Stein61,Pipkin86,Steigmann90} is that if the bending modulus is sufficiently small the stress field approaches a well-defined compression-free profile (see {\emph{e.g.}} fig.~\ref{fig:Profiles}), which is {\emph{independent}} on the bending modulus. 
Refs.~\cite{Davidovitch11,Davidovitch12,King12,Hohlfeld15,Taffetani17} describe the energetic hierarchy in this limit (and consequent force balance) as a ``far from threshold'' expansion in the inverse bendability parameter, $\epscap$ (rather than the familiar post-buckling approach, which is a Landau-type expansion in the wrinkle amplitude, around the compressed (unstable) planar state solution). The crucial point is that the energetic hierarchy (as well as the stress field) retain a structure similar to (\ref{eq:UHeirarchy}), whereby the bending energy is sub-dominant ({\emph{i.e.}} $\Delta \Ubend/\Delta \Ustrain \to 0$ as $\epscap \to 0$). Hence, although the presence of a small bending rigidity has in this case an {\emph{implicit}} effect on the mechanics, enabling the formation of a compression-free stress field (and correspondingly lower values of $\Delta \Usurf + \Delta \Ustrain$ than the unwrinkled, compressed state \cite{Davidovitch11}) any {\emph{explicit}} effect of the bending rigidity on the energy or stress field is perturbative, and can be ignored, as long as one is careful to properly employ tension field theory (rather than membrane theory) in solving the FvK equations \eqref{eqn:FvK2-new-in},\eqref{eqn:FvK2-new-out} (as described in Appendix A).

\section{On the limit of vanishing thickness}

Nadermann \emph{et al.}~\cite{Nadermann13} proposed a model to support their reliance on ``.. the intuitive assumption that the stretch contributions vanish as the film's thickness reduces to zero". Furthermore, they made a constitutive assumption that the tension $\Ti$ is linear in sheet thickness, with intercept $\Ti=2\gsv$ at $t=0$. Notwithstanding some differences, their model (page 2 of Supplementary Information (SI) \cite{Nadermann13}) is conceptually similar to our analysis of the FvK equations in \S\ref{sec:details} (and in refs.~\cite{Vella10,Schroll13}), exploiting axial symmetry and neglecting any explicit dependence of the stress on the bending modulus. Since we showed in \S\ref{sec:qualitative} that the characteristic capillary-induced stress, $T^\ast$, and its non-perturbative effect can be understood conceptually through energetic scaling arguments ({\emph{i.e.}}~without even needing to solve the FvK equations), a careful reader may wonder which of the assumptions and approximations of Nadermann \emph{et al.}~\cite{Nadermann13} led to 
such different conclusions to those presented here. We thus include here a summary of the key differences between the analysis of Nadermann \emph{et al.}~\cite{Nadermann13} and ours. Each bulleted point explains a difference, and assesses its implications; note that the discussion of the third point parallels the energetic considerations in \S2.1, as well as \S3.3.
\\

$\bullet$ {\emph{Spherical cap:}} In FvK-based theory, one obtains Eqs.(\ref{eqn:FvK2-new-in},\ref{eqn:FvK2-new-out}) -- a coupled set of nonlinear ODEs for the deflection, $\zeta(r)$, and the stress potential, $\psi(r)$, subject to appropriate BCs (\S3.2 and Appendix \ref{sec:BCs}). These equations are solved numerically, yielding the radial profiles of the deflection (fig.~\ref{fig:Shapes}) and stress (fig.~\ref{fig:Profiles}). In contrast, Nadermann \emph{et al.}~\cite{Nadermann13} assume that the deflection can be approximated by a spherical cap, simplifying considerably the analysis. 

\emph{Significance:} As we discussed in \S5.3.2, our solution shows that, unless the system is deeply in the non-perturbative regime ({\emph{i.e.}} pre-tension much smaller than capillary-induced stress, $\tauinf \ll 1$), the spherical cap assumption is well justified. 
\\

$\bullet$ {\emph{Membrane theory versus tension-field:}} As explained in detail elsewhere (see Schroll \emph{et al.}~\cite{Schroll13} as well as Appendix B), neglecting an explicit dependence of the stress on bending modulus (in the limit of high capillary bendabilty, Eq.~(\ref{eq:define-bendability}), akin to Eq.~S3 in Nadermann \emph{et al.}~\cite{Nadermann13}), must be done with care. A membrane theory solution (akin to Vella \emph{et al.}~\cite{Vella10}), such as that sought by Nadermann \emph{et al.}~\cite{Nadermann13} (having argued that the effect of the bending force in the boundary layer is negligible, see page 1 of supplementary information \cite{Nadermann13}), is valid  only if the result is a purely tensile stress. A (partially) compressive solution signals that the actual stress profile is described by tension field theory, yielding an asymptotic compression-free stress field (fig.~\ref{fig:Profiles}); this stress state is markedly different from the membrane theory solution. 

\emph{Significance:} Our energetic scaling analysis in \S2, which is indifferent to the exact stress field in the sheet, indicates that the conceptual distinction between perturbative and non-perturbative effects of the capillary-induced stress, $T^\ast\sim \glv^{2/3}Y^{1/3}$, can be realized also by a membrane theory calculation. Indeed, while the tension-field calculation is crucial for making any  quantitative predictions, the mere distinction between perturbative and non-perturbative effects of the capillary-induced stress can be realized also by a membrane theory calculation; such an (albeit unstable) solution would yield a plateau $\Ti/\glv \sim  T^\ast$ as $\Tinf/\glv \to 0$, rather than the slow logarithmic decay observed in fig.~\ref{fig:FiniteTgamma}. \\

$\bullet$ {\emph{The limit of ``vanishing thickness'':}} Carrying out a membrane theory-like calculation, Nadermann \emph{et al.}~\cite{Nadermann13} obtain an equation (Eq.~S13 of ref.~\cite{Nadermann13}) that expresses the stress in the vicinity of the contact line as a sum of pre-tension ($\Tinf$ in our notation, or surface energy terms in the interpretation of \cite{Nadermann13}), and another contribution due to the stretching induced by the drop. This second contribution is expressed (using the notations of our paper) as a product of the stretching modulus, $Y$, and the terms, $\tfrac{\rmur(\Rin)}{\Rin} ,\sin^2\phi$, reflecting, respectively, contributions to radial strain due to in-plane and out-of-plane displacements. Arguing that these strain terms  ``remain bounded'', Nadermann \emph{et al.}~\cite{Nadermann13} conclude that, upon multiplying by $Y \sim Et$, the capillary-induced contributions can be ignored in the limit of ``vanishing thickness" $t \to 0$.

\emph{Significance:} As we emphasized (see specifically \S5.1), a description using membrane/tension-field theory ({\emph{i.e.}} the FvK equations with no explicit dependence on bending modulus) is valid only in an {\emph{intermediate}} parameter range, which {\emph{does not}} include the regime $t \ll \glv/E$ (where capillary-induced tensile strains imply a highly non-Hookean response). Hence, one should be careful {\emph{not to}} consider the unconditional limit $t\to 0$, but rather use the two dimensionless parameters that involve the sheet's thickness ($\epscap \ll 1$ and $\tglv \ll 1$, Eqs.~(\ref{eq:define-bendability},\ref{eq:DGSchroll})), such that the system remains in the parameter regime defined by the {\emph{two}} inequalities
(\ref{eqn:asylimit}). In this parameter regime (which we showed in \S5.1 to characterize essentially all experimental systems in Refs.~\cite{Nadermann13,Schulman15,Huang07,Schroll13,Toga13}), the mechanics is governed by $\btauinf$ (\ref{eq:define-tauinf-2}), namely, the ratio between the pre-tension, $\Tinf$, and the characteristic capillary-induced stress, $T^\ast \sim \glv^{2/3}Y^{1/3}$. As our energetic arguments in \S2 already showed (and our quantitative solution of the FvK equations confirmed) 
capillary-induced terms (specifically, $Y\phi^2$) may be larger or smaller than the pre-tension (or any scale proportional to surface energy), depending on the value of $\btauinf$. Overlooking this subtle, intermediate-asymptotic nature of the ``vanishing thickness" limit, seems to underlie the conclusion of Nadermann \emph{et al.}~\cite{Nadermann13} that capillary-induced stretching can be ignored in analyzing their data.

\end{appendix}


\end{document}